\documentclass[showpacs,preprintnumbers,amsmath,amssymb,aps,prc,floatfix]{revtex4-1}

\usepackage[usenames,dvipsnames]{xcolor} 
\usepackage{graphicx}
\usepackage{dcolumn}
\usepackage{bm}
\usepackage{float}
\usepackage{dcolumn}
\usepackage{dsfont}
\usepackage{todonotes}
\usepackage{multirow}
\begin{document}
\title{Final-state interactions in two-nucleon knockout reactions}
\author{Camille Colle}
\email{Camille.Colle@UGent.be}
\author{Wim Cosyn}
\email{Wim.Cosyn@UGent.be}
\author{Jan Ryckebusch}
\email{Jan.Ryckebusch@UGent.be}
\affiliation{Department of Physics and Astronomy,\\
 Ghent University, Proeftuinstraat 86, B-9000 Gent, Belgium}
\date{\today}
\begin{abstract}
\begin{description}
\item[Background]

Exclusive two-nucleon knockout after electroexcitation of nuclei
($A(e,e'NN)$ in brief) is considered to be a primary source of
information about short-range correlations (SRC) in nuclei. For a
proper interpretation of the data, final-state interactions (FSI) need
to be theoretically controlled.
\item[Purpose] Our goal is to quantify the role of FSI effects in
  exclusive $A(e,e'pN)$ reactions for four target nuclei representative for the
  whole mass region. Our focus is on processes that are SRC driven.
  We investigate the role of FSI for two characteristic detector setups
  corresponding with a ``small'' and ``large'' coverage of the available phase
  space.

\item[Method] Use is made of a factorized expression for the
  $A(e,e'pN)$ cross section that is proportional to the two-body
  center-of-mass (c.m.) momentum distribution of close-proximity
  pairs. The $A(e,e'pp)$ and $A(e,e'pn)$ reactions for the target
  nuclei $^{12}$C, $^{27}$Al, $^{56}$Fe and $^{208}$Pb are
  investigated.  The elastic attenuation mechanisms in the FSI are
  included using the relativistic multiple-scattering Glauber
  approximation (RMSGA). Single-charge exchange (SCX) reactions are
  also included. We introduce the nuclear transparency $T^{pN}_{A}$,
  defined as the ratio of exclusive $(e,e'pN)$ cross sections on
  nuclei to those on ``free'' nucleon pairs, as a measure for the
  aggregated effect of FSI in $pN$ knockout reactions from nucleus
  $A$. A toy model is introduced in order to gain a better understanding of the $A$ dependence of $T_{A}^{pN}$.

\item[Results] The transparency $T^{pN}_{A}$ drops from $ 0.2-0.3 $
for $^{12}$C to $0.04-0.07$ for $^{208}$Pb. For all considered
kinematics, the mass dependence of the $T^{pN}_{A}$ can be captured
by the power law $T^{pN}_{A} \propto A^{- \lambda}$ with $ 0.4
\lesssim \lambda \lesssim 0.5 $. Apart from an overall reduction
factor, we find that FSI only modestly affects the distinct features
of SRC-driven $A(e,e'pN)$ which are dictated by the c.m. distribution
of close-proximity pairs.
\item[Conclusion] The SCX mechanisms represent a relatively small (order of a few percent) contribution of SRC-driven $A(e,e'pN)$ processes. The mass dependence of FSI effects in exclusive $A(e,e'pN)$ can be captured in a robust power law and is in agreement with the predictions obtained in a toy model.
\end{description}
\end{abstract}
\maketitle
\section{Introduction}
\label{sec:intro}

Nuclear SRC are an essential ingredient of the dynamics of nuclei at
large momenta and energies. The short- and medium-range components of
the nucleon-nucleon interaction induce beyond mean-field high-momentum
and high-density fluctuations in the nuclear medium, thereby giving
rise to fat tails in the nuclear momentum distributions
\cite{Wiringa:2013ala,Rios:2013zqa,Vanhalst:2014cqa}.  The magnitude
of nuclear SRC has been linked to plateaus in ratios of cross sections
of inclusive electron scattering off different nuclei
\cite{Egiyan:2003vg,Egiyan:2005hs,Fomin:2011ng}, and to the size of
the EMC effect \cite{Hen:2013oha}.

Nuclear SRC can be studied in exclusive two-nucleon knockout processes
with hadronic and electroweak probes.  In appropriately selected
kinematics, those reactions give access to the dynamics and isospin composition of the initial
nucleon pair.  In the 1990s, high-resolution $A(e,e'pp)$ measurements
carried out at MAMI \cite{Rosner200099,Blomqvist:1998gq} and NIKHEF
\cite{Onderwater:1997zz,Onderwater:1998zz,Starink2000} could determine
the transition to a specific final state of the residual $A-2$
nucleus. When comparing to data for the $^{16}$O$(e,e'pp)$ transition
to the $0^+$ ground state of $^{14}$C, model calculations
\cite{Ryckebusch:2003tu,Ryckebusch:1997gn,Giusti:1997pa} showed the
clear dominance of SRC contributions to the cross section at low
c.m.~pair momentum, where the initial pair is in a relative $S$-state.
The EVA collaboration at Brookhaven National Laboratory (BNL) measured
the $^{12}$C$(p,ppn)$ reaction \cite{Tang:2002ww} as a function of the
initial neutron momentum.  For neutron momenta above the Fermi surface
($\sim 220$~MeV) the data showed a clear angular correlation between
the initial proton and neutron momenta with backward angles
($>90^{\circ}$) dominating.  For momenta below the Fermi surface the
angular correlation between the two nucleon momenta is almost random.
This picture was later confirmed by a $^{3}$He$(e,e'pp)$ experiment
performed in Jefferson Lab \cite{Niyazov:2003zr}. More recently,
$^{12}$C$(e,e'pN)$ \cite{Shneor:2007tu,Subedi:2008zz} and
$^{4}$He$(e,e'pN)$ \cite{Korover:2014dma} measurements (both at
Jefferson Lab) provided proof that in the probed kinematics about 20\%
of the nucleons in nuclei form correlated pairs.  Of those, about 90\%
is of the proton-neutron type \cite{Hen31102014}, illustrating the
dominance of tensor correlations in the nucleon momentum region of
300-500~MeV/c.  A feature that emerges from all those experimental
investigations, is that SRC pairs are mostly in a back-to-back
configuration with a high relative and small c.m.~momentum, whereby
small and large are defined relative to the Fermi momentum.

In this paper we focus on the effect of FSI in SRC-driven
high-virtuality $A(e,e'pN)$ cross sections.  In Sec.~\ref{sec:model},
we discuss the approximations underlying the factorized form of the
$A(e,e'pN)$ cross section (detailed in Ref.~\cite{Colle:2014}) and how
we implement the FSI.  Using the factorized $A(e,e'pN)$ cross-section
expression, we show in Sec.~\ref{sec:ratios} that cross-section ratios
can be directly related to the ratios of the integrated distorted
two-body c.m.~momentum distributions of close-proximity nucleon pairs.
In Sec.~\ref{sec:results}, we apply the developed model to four
different target nuclei ($^{12}$C, $^{27}$Al, $^{56}$Fe, $^{208}$Pb)
and two very different kinematics probing SRC pairs.  First, the
kinematics of the $A(e,e'pp)$ cross-section measurements with the
CEBAF Large Acceptance Spectrometer (CLAS) \cite{Hen31102014} covering
a very large phase space. Second, the kinematics of an experimental
setup with a very restricted phase-space coverage
\cite{Shneor:2007tu}.  We extract the nuclear transparencies for
two-nucleon knockout and compare them to single-particle knockout
transparencies extracted from $A(e,e'p)$ measurements. We propose
parameterizations for the mass dependence of the $A(e,e'pN)$
transparencies in the form of a power law and study its robustness.
The opening-angle distribution for the initial correlated nucleon pair
is shown to be dominated by backward angles, with little modification
after the inclusion of FSI. A toy model that captures the essential
features of elastic attenuation mechanisms in $A(e,e'NN)$ is
proposed. This toy model allows us to gain a more qualitative understanding
of the mass dependence of the nuclear transparency.
Conclusions are given in Sec.~\ref{sec:conclusion}.

%
\section{Model}
\label{sec:model}
\subsection{Factorization of the $A(e,e'pN)$ cross section}
\label{sec:factorization}

\begin{figure}
\centering
\includegraphics[width=\textwidth,viewport= 0 330 1200 600,clip]
{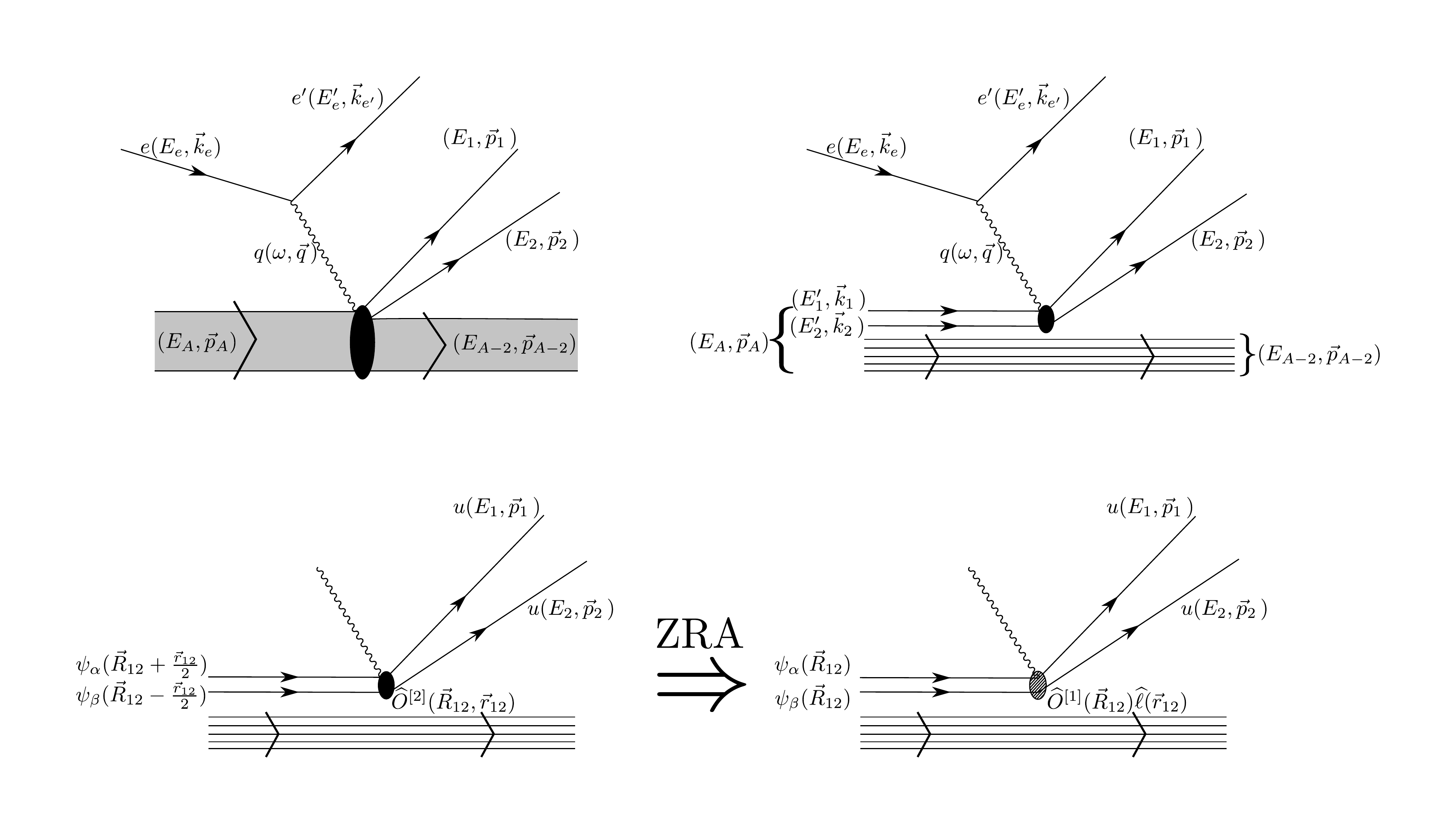}
\caption{(left) Sketch of the exclusive $A(e,e'pN)$ reaction with all
  kinematic variables. (right) The $A(e,e'pN)$ reaction in the
  impulse- and spectator approximation.}
\label{fig:twoNNknockout_diag}
\end{figure}

We consider exclusive electroinduced knockout of a correlated
proton-nucleon ($pN$) pair from the target nucleus $A$
\begin{align}
	e + A \rightarrow e' + (A-2)^{*} + p + N\,.
\end{align}
In this paper we solely deal with reactions whereby the residual
$(A-2)^{*}$ is left with little or no excitation energy. This
condition is essential for keeping the number of contributing reaction
mechanisms under control.  

Let ($\vec{k}_1, \vec{k}_2$) and ($\vec{p}_1, \vec{p}_2$) be the
initial and final three-momenta of the nucleon pair.  We label the struck
proton with ``1'' and the recoiling nucleon with ``2''.  In the
impulse approximation, in which the exchanged momentum is absorbed by
a single nucleon, we have that $\vec{p}_1 = \vec{k}_1 + \vec{q}$ with
$\vec{q}$ the transferred three-momentum of the virtual photon
(Fig.~\ref{fig:twoNNknockout_diag}). We define the c.m.~$\vec{P}_{12}$
and relative momentum $\vec{k}_{12}$ of the initial pair as,
\begin{align}
	\vec{P}_{12} = \vec{k}_{1} + \vec{k}_{2}, \hspace{0.05\textwidth}
	\vec{k}_{12} &= \frac{ \vec{k}_1 - \vec{k}_2 }{2} \; . 
\end{align}
The corresponding c.m.~and relative coordinates are denoted by
$\vec{R}_{12}$ and $\vec{r}_{12}$.

By selecting events with a large $|\vec{q}\,|$ (large in comparison to
the initial momenta ($\vec{k}_1, \vec{k}_2$) of the nucleon pair) and
requiring that one of the measured nucleons carries a significant
fraction of the exchanged momentum $|\vec{q}\,|$, the contribution
from the exchange term in which nucleon ``2'' absorbs the photon can
be made negligible. Indeed, above the Fermi momentum, the
$\vec{k}_{12}$ distribution of the pairs is strongly decreasing with
increasing $|\vec{k}_{12}|$~\cite{Colle:2014,Vanhalst:2014zt}. This makes it highly unlikely
that the fast nucleon in the final state is not the one that absorbed
the virtual photon.

\begin{figure}
\centering
\includegraphics[width=\textwidth,viewport= 0 0 1000 300,clip]
{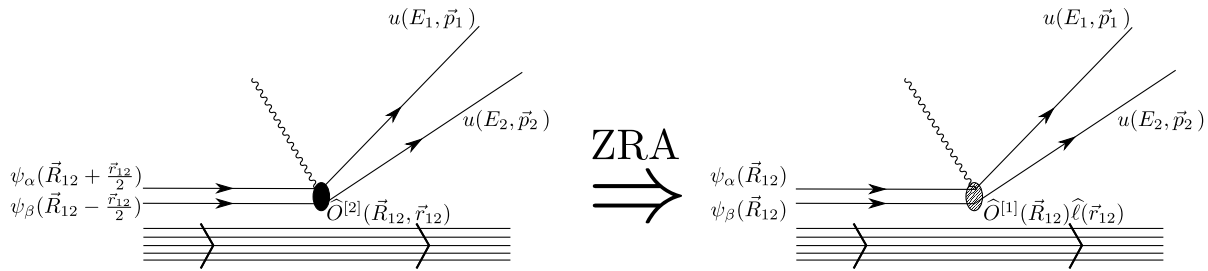}
\caption{A sketch of the zero-range approximation (ZRA) which underlies the
  factorized expression of the exclusive $A(e,e'pN)$ cross section. }
\label{fig:twoNNknockout_diag2}
\end{figure}

As outlined in Refs.~\cite{Ryckebusch:1996wc,Colle:2014}, in
kinematics probing SRC pairs, it is possible to factorize the
$A(e,e'pN)$ cross section in a product of a function depending on the
relative momentum $\vec{k}_{12}$, and a part depending on the
c.m.~momentum $\vec{P}_{12}$ of the initial $pN$ pair
\begin{equation}
\frac{\textrm{d} ^{8} \sigma (e,e'pN)}{\textrm{d}^{2} \Omega_{k_{e'}}
  \textrm{d}^{3}\vec{p}_{1} \textrm{d}^{3}\vec{p}_{2}} = \frac{M_A M_{A-2}}{E_A E_{A-2}} \frac{1}{(2\pi)^{3}} f_{\text{rec}} \sigma_{epN}(\vec{k}_{12}) F_{A}^{pN,D}(\vec{P}_{12}) \,,
\label{eq:eeNNfactorized}
\end{equation}
 with $\Omega_{k_{e'}}$ the solid
angle of the scattered electron, $f_{\text{rec}}$ the recoil factor, 
\begin{align}
f_{\text{rec}} = \frac{ \left| 1 - \frac{E_{e'}}{ E_{2}} \frac{ \vec{p}_{2} \cdot \vec{k}_{e'}}{| \vec{k}_{e'} |^2}  \right| }{ \left| 1 + \frac{E_{e'}}{E_{A-2}}  \frac{\vec{P} \cdot \vec{k}_{e'}}{| \vec{k}_{e'} |^2} \right|} \, ,
\end{align}
and $M_{A}$, $E_{A}$ ($M_{A-2}$,$E_{A-2}$) the rest mass and energy of the initial (recoiling $A-2$) nucleus.
$\sigma _{epN}(\vec{k}_{12})$ encodes the virtual-photon coupling to
a correlated $pN$ pair with relative momentum $\vec{k}_{12}$.
$F_{A}^{pN,D}(\vec{P}_{12})$ is the distorted c.m.~momentum
distribution of the close-proximity pair that absorbs the photon.  The
factorized cross-section expression of Eq.~(\ref{eq:eeNNfactorized})
hinges on the validity of the zero-range approximation (ZRA), which
amounts to putting the relative pair coordinate $\vec{r}_{12}$ to zero
(Fig.~\ref{fig:twoNNknockout_diag2}). Thereby, the amplitude for
photo-absorption on a close-proximity pair that involves the product
of two IPM wave functions $\psi_{\alpha}(\vec{R}_{12} +
\frac{\vec{r}_{12}}{2} )$ and $\psi_{\beta}(\vec{R}_{12} -
\frac{\vec{r}_{12}}{2} )$ and a two-body operator
$\widehat{O}^{[2]}(\vec{R}_{12},\vec{r}_{12})$ (left panel of
Fig.~\ref{fig:twoNNknockout_diag2}) is written as a product of a
one-body operator $\widehat{O}^{[1]}$ evaluated at the c.m.~coordinate
$\vec{R}_{12}$ and a correlation operator $\widehat{\ell}$ that
depends only on the relative coordinate $\vec{r}_{12}$ (right panel of
Fig.~\ref{fig:twoNNknockout_diag2}). In nuclei, $\widehat{\ell}$
has a complicated spin and isospin structure. The ZRA acts as a
projection operator on the short-range components of
the wave function corresponding with the relative motion of the pair.
Throughout this paper the factorized cross section of
Eq.~(\ref{eq:eeNNfactorized}) is used.
The validity of this expression~(\ref{eq:eeNNfactorized}) has been experimentally
verified. The proposed factorization of the cross section in terms of
$F_{A}^{pp,D}$ was first confirmed in $^{12}$C$(e,e'pp)$ measurements
back in the late 1990s~\cite{Blomqvist:1998gq}. An effort is on its way
to extract the width of the $F_{A}^{pp,D}$ distribution in $A(e,e'pp)$
measurements on $^{12}$C, $^{27}$Al, $^{56}$Fe and $^{208}$Pb~\cite{Hen:2012jn,OrPriv}, and compare them with the theoretical predictions~\cite{Colle:2014}.
Another striking prediction of the expression~(\ref{eq:eeNNfactorized}) is that the $A(e,e'pN)$ cross section is proportional to the number of close-proximity $pN$ pairs in the target nucleus. As a result, it can be inferred that the $A$ dependence of the $A(e,e'pp)$ cross section is much softer than naive FSI-corrected $Z(Z-1)/2$ counting. Recent measurements of the $A(e,e'pp)/^{12}\text{C}(e,e'pp)$ ratios are completely in line with those predictions~\cite{Colle:massdep}. The measured and predicted $^{208}$Pb/$^{12}$C $(e,e'pp)$ cross section ratio, for example, is a mere five whereas the naive prediction is over two hundred.

%

%
%
\subsection{Final-state interactions}
\label{sec:final-state-interactions}
We include two FSI mechanisms in our model. First, attenuation (ATT) of the
outgoing nucleons upon traversing the recoiling nucleus. Second,
single-charge exchange (SCX), i.e.~an outgoing proton (neutron)
rescattering into a neutron (proton).  The attenuation effect is
calculated in the relativistic multiple-scattering Glauber
approximation (RMSGA)~\cite{Ryckebusch:2003fc,Cosyn:2013qe}.  The
RMSGA is based on high-energy diffractive scattering. It uses an
eikonal form for the rescattering amplitude dominated by the central
term, neglecting spin-dependent attenuation. The RMSGA is fully parameterized
in terms of nucleon-nucleon scattering data.  We systematically use
``FSI'' upon referring to the combined effect of attenuation and single-charge exchange.
Throughout this paper, we refer to $A(e,e'pN)$ results that ignore the effect of FSI as ``ZRA'' results. 

The distorted c.m.~momentum distribution $F_{A}^{pN,D}(\vec{P}_{12})$
in Eq.~(\ref{eq:eeNNfactorized}) is defined in the following way
\begin{multline}\label{eq:FD_RMSGA}
	F_{A}^{pN,D}(\vec{P}_{12}) = \sum_{\alpha, \beta} F_{A}^{pN,D;\alpha \beta}(\vec{P}_{12}) 
	=\sum_{\substack{ s_1,s_2 \\ \alpha \beta}}
\left| \int d\vec{R}_{12} \, e^{i  \vec{P}_{12}
\cdot \vec{R}_{12} } \bar{u}(\vec{k}_1,s_1) \psi_{\alpha}(\vec{R}_{12}) \bar{u}(\vec{k}_2,s_2) \psi_{\beta}(\vec{R}_{12}) \mathcal{F}_{\text{RMSGA}}(\vec{R}_{12})
\right|^{2} \, .
\end{multline}
Here, $u(\vec{k},s)$ is the positive-energy free Dirac spinor and $
\left( \psi_{\alpha},\,\psi_{\beta} \right)$ are relativistic
mean-field wave functions with IPM quantum numbers $ \left(
\alpha,\beta \right)$ computed in the Serot-Walecka model
\cite{Furnstahl:1996wv}. The contribution from a specific IPM nucleon
pair with quantum numbers $\left( \alpha,\beta \right)$ is denoted as
$F_{A}^{pN,D;\alpha \beta}(\vec{P}_{12})$.  We consider experimental
conditions whereby the precise state of the residual $A-2$ nucleus is
not resolved. As a result, the sum over $ \left( \alpha,\beta \right)$
extends over all occupied $pN$ pairs.  In the practical implementation
of Eq.~(\ref{eq:FD_RMSGA}), we neglect the projection on the lower
components of the plane-wave Dirac spinors.  The rescattering of the
ejected pair with the remaining $A-2$ spectators, encoded in the
standard Glauber phase $\mathcal{F}_{\text{RMSGA}}$, is computed in
the RMSGA~\cite{Ryckebusch:2003fc}.  The interactions among the two
ejected nucleons do not affect $F^{pN,D}_{A}(\vec{P}_{12})$ due to
c.m.~momentum conservation, and they are effectively included in
$\sigma_{epN}(\vec{k}_{12})$. The implementation of reinteractions
between the ejected nucleons is not addressed in this article, as
$\sigma_{epN}(\vec{k}_{12})$ drops out in the cross section ratios
defined in Sec.~\ref{sec:ratios}.

%
%
The SCX mechanisms are treated in a semi-classical manner. The joint
probability for the struck proton, labelled ``1'', undergoing SCX while
the recoiling nucleon of the SRC pair, labelled ``2'', is not
undergoing any SCX, is given by
\begin{align}
P_{\textrm{CX,A}}^{\mathbf{[1]}2, pN } = 
	\sum_{\alpha,\beta} \frac{\int \textrm{d}^3 \vec{R}_{12} \, 
P^{[\boldsymbol \alpha] \beta}_{\textrm{CX}}(\vec{R}_{12}) \left[ 1 - 
P^{ \alpha [\boldsymbol \beta]}_{\textrm{CX}}(\vec{R}_{12})\right] F^{pN,D; 
\alpha \beta}_{A}(\vec{R}_{12})}{\int \textrm{d}^3 \vec{R}_{12} \, 
F^{pN,D;\alpha \beta}_{A}(\vec{R}_{12})} \, .
    \label{eq:cx_prob_example}
\end{align}
As in Eq.~(\ref{eq:FD_RMSGA}), the sum over $ \left( \alpha,\beta
\right)$ extends over all the occupied $pN$ pairs. Further, the square
bracket $\mathbf{[1]}$ identifies the nucleon subject to SCX.
In Eq.~(\ref{eq:cx_prob_example}), the probability that an initial
nucleon with quantum numbers $\alpha$ (with correlated partner with
quantum numbers $\beta$) has undergone a SCX after a hard interaction
at c.m.~coordinate $\vec{R}_{12}$ is given by $P^{[\boldsymbol \alpha]
  \beta}_{\textrm{CX}}(\vec{R}_{12})$, and is weighted with the
RMSGA-corrected probability $F^{pN,D;\alpha \beta}_{A}(\vec{R}_{12})$
of finding the two nucleons at c.m.~coordinate $\vec{R}_{12}$.
Similar expressions to Eq.~(\ref{eq:cx_prob_example}) can be written
for the situations where only the recoil nucleon ``2'' is subject to
SCX ($P_{\textrm{CX,A}}^{1[\mathbf{2}],pN}$), both nucleons in the
pair are subject to SCX ($P_{\textrm{CX,A}}^{[\mathbf{12}],pN}$) or
none of the nucleons in the pair are subject to SCX
($P_{\textrm{CX,A}}^{12,pN}$). In those situations, the factor
$P^{[\boldsymbol \alpha] \beta}_{\textrm{CX}}(\vec{R}_{12}) \left[ 1 -
  P^{ \alpha [\boldsymbol \beta]}_{\textrm{CX}}(\vec{R}_{12})\right]$
in the numerator of Eq.~(\ref{eq:cx_prob_example}), is replaced by
respectively the factor $\left[1-P^{[\boldsymbol \alpha]
    \beta}_{\textrm{CX}}(\vec{R}_{12})\right] P^{\alpha[\boldsymbol \beta
  ]}_{\textrm{CX}}({R}_{12})$, $P^{[\boldsymbol \alpha]
  \beta}_{\textrm{CX}}(\vec{R}_{12}) P^{\alpha[\boldsymbol \beta
  ]}_{\textrm{CX}}({R}_{12})$, and $[1-P^{[\boldsymbol \alpha]
    \beta}_{\textrm{CX}}(\vec{R}_{12}) ][1-P^{
    \alpha[\boldsymbol \beta]}_{\textrm{CX}}(\vec{R}_{12})]$. One has
$P_{\textrm{CX,A}}^{[\mathbf{1}]2, pN} +
P_{\textrm{CX,A}}^{1[\mathbf{2}], pN} +
P_{\textrm{CX,A}}^{[\mathbf{12}], pN}+ P_{\textrm{CX,A}}^{12, pN} =
1$.

The SCX probabilities $P^{[\boldsymbol \alpha] 
\beta}_{\textrm{CX}}(\vec{R}_{12})$ are calculated in a semi-classical 
approximation. Thereby, the probability of charge-exchange rescattering for a 
nucleon with bound-state IPM quantum numbers $\alpha$ that is brought in a 
continuum state at the coordinate $\vec{r}$ is modelled by
\begin{align}
	 P_{\textrm{CX}}^{[\boldsymbol \alpha] \beta}(\vec{r} \, ) = 1 - 
\exp\left[ - \sigma_{\textrm{CX}}(s) \;  \int_{z}^{+\infty} \textrm{d}z' 
\rho^{\alpha \beta}_{A-2}(z')\right] \, .
     \label{eq:CX_prob_r}
\end{align}
The $z$-axis is chosen along the direction of propagation of the nucleon 
undergoing SCX ($[\boldsymbol \alpha]$). The $\rho^{\alpha \beta}_{A-2}$ is the one-body density of the 
recoiling $A-2$ nucleus that contributes to the SCX reaction. For an ejected proton (neutron) 
only the neutron (proton) density of the recoiling nucleus affects SCX 
reactions. 
The parameter $\sigma_{\textrm{CX}}(s)$ in Eq.~(\ref{eq:CX_prob_r}) can be 
extracted from elastic proton-neutron scattering data~\cite{PhysRevC.30.566}, 
with $s$ the total c.m.~energy squared of the two nucleons involved in the SCX. 
In  Ref.~\cite{PhysRevC.50.2742}, it was shown that $\sigma_{\textrm{CX}}(s)$ 
obeys the relation
\begin{align}
	\sigma_{\textrm{CX}}(s) = 0.424 \frac{s}{s_{800}} \textrm{ fm}^{2} \, ,
	\label{eq:cx_cross_param}
\end{align}
where $s_{800}$ is the c.m.~energy squared for a collision between a neutron 
with 800~MeV kinetic energy and a stationary proton.
The value $0.424$ fm$^{2}$ is obtained by integrating the elastic 
$pn$ differential cross section with $s$=$s_{800}$ at backward scattering angles 
dominated by charge-exchange~\cite{PhysRevC.30.566}. 
The parameterization of Eq.~(\ref{eq:cx_cross_param}) is valid for lab frame 
momenta in the interval $ [0.1,100]$ GeV/c.

In Eq.~(\ref{eq:cx_prob_example}), the weight factor $F^{pN,D;\alpha 
\beta}_{\text{A}}(\vec{R}_{12})$ gives the attenuation corrected probability to find a pair 
$ \left( \alpha,\beta \right)$ at a coordinate $\vec{R}_{12}$ 
\begin{align}
	F^{pN,D; \alpha \beta}_{A}(\vec{R}_{12}) = 
\lim_{\vec{r}_{12} \rightarrow \vec{0}} \, | \psi_{\alpha}(\vec{R}_{12} + 
\frac{\vec{r}_{12}}{2}) |^{2} | \psi_{\beta}(\vec{R}_{12} - 
\frac{\vec{r}_{12}}{2}) |^{2} | 
\mathcal{F}_{\textrm{RMSGA}}(\vec{R}_{12} \pm \frac{\vec{r}_{12}}{2}) |^{2}  \, .
    \label{eq:CX_cond_dens}
\end{align}
 Note that $F^{pN,D;\alpha \beta}_{A}(\vec{R}_{12})$ is the Fourier
 transform of $F_{A}^{pN,D;\alpha \beta}(\vec{P}_{12})$ appearing in
 Eq.~(\ref{eq:eeNNfactorized}). In the limit of vanishing FSI
 ($\mathcal{F}_{\textrm{RMSGA}} \equiv 1$) Eq.~(\ref{eq:CX_cond_dens})
 reduces to the probability of finding two IPM nucleons at the same
 coordinate $\vec{R}_{12}$.

\begin{figure}
\includegraphics[width=0.45\textwidth]{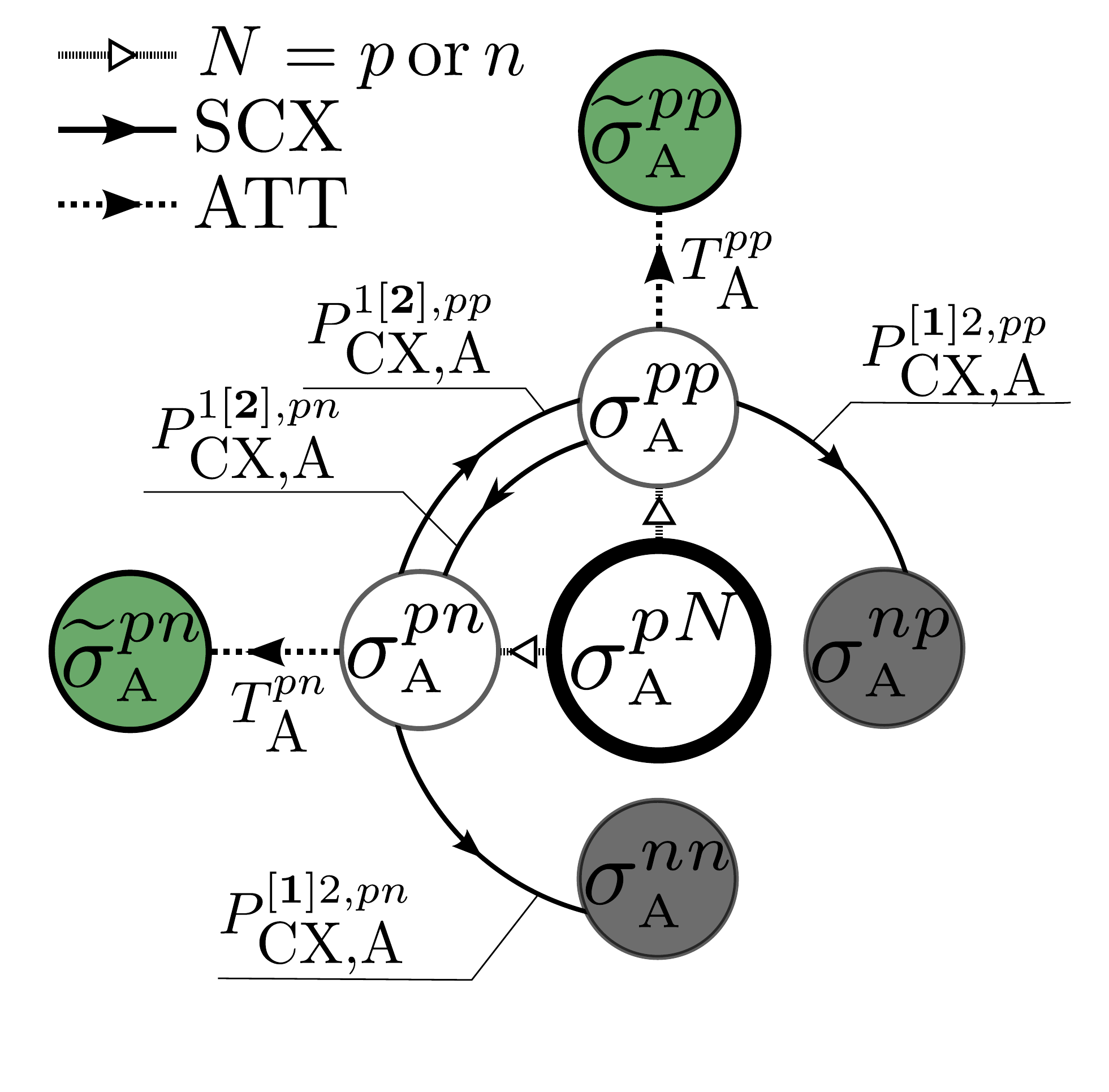}
\caption{(Color online) A flow diagram illustrating the different FSI effects
  included in our model calculations. The center of the diagram
  ($\sigma^{pN}_{\textrm{{\tiny A}}}$) denotes the plane-wave ZRA
  $A(e,e'pN)$ cross section.  The
  $\widetilde{\sigma}^{pp}_{\textrm{{\tiny A}}}$ and
  $\widetilde{\sigma}^{pn}_{\textrm{{\tiny A}}}$ correspond with the
  observed $A(e,e'pp)$ and $A(e,e'pn)$ cross sections.  The solid
  arrows denote SCX reactions. The dashed arrows denote the 
  attenuation (ATT).}
\label{fig:SCX_RMSGA_diag}
\end{figure}

The flow diagram in Fig.~\ref{fig:SCX_RMSGA_diag} shows an overview of the 
different FSI mechanisms that are included in the $A(e,e'pN)$ reactions 
considered in this article. The initial plane-wave (vanishing FSI) ZRA 
$A(e,e'pN)$ cross section ($\sigma^{pN}_{A}$) is positioned at the center.
The partner nucleon can be a 
proton ($\sigma^{pp}_{A}$) or a neutron ($\sigma^{pn}_{A}$). The observable 
cross sections are denoted with $\widetilde{\sigma}^{pp}_{A}$ and 
$\widetilde{\sigma}^{pn}_{A}$. The sources and sinks between the 
different $\widetilde{\sigma}^{NN}_{A}$ through the SCX mechanism are denoted 
with the solid arrows. The dashed arrows denote the RMSGA attenuation 
contribution, quantified by means of the nuclear transparency $T^{pN}_{A}$,
defined as the ratio of the $A(e,e'pN)$ cross section with and without the 
RMSGA attenuations (see Eq.~(\ref{eq:transparency})). It is a 
measure for attenuation caused by the nuclear medium. 
The different contributions to the final cross sections
$\widetilde{\sigma}^{pp}_{A}$, $\widetilde{\sigma}^{pn}_{A}$ can be
visually deduced by following all possible paths from
$\sigma^{pN}_{A}$ to $\widetilde{\sigma}^{pp}_{A}$ or
$\widetilde{\sigma}^{pn}_{A}$ in
Fig.~(\ref{fig:SCX_RMSGA_diag}). Because we only account for
single-charge exchange, the SCX arrows can only be used at most once
for each particle in every path, meaning
$P_{\textrm{CX,A}}^{1[\mathbf{2}],pN}
P_{\textrm{CX,A}}^{[\mathbf{1}]2,pN}$ is allowed but
$P_{\textrm{CX,A}}^{[\mathbf{1}]2,pN}
P_{\textrm{CX,A}}^{[\mathbf{1}]2,pN}$ or
$P_{\textrm{CX,A}}^{1[\mathbf{2}],pN}
P_{\textrm{CX,A}}^{1[\mathbf{2}],pN}$ are not.  
The missing SCX arrows (from and between $\sigma_{A}^{nn}$,
$\sigma_{A}^{np}$) are neglected as we assume that the struck nucleon
is a proton.  We argue that this is a valid approximation. First, the
photon-neutron coupling is a fraction of the photon-proton
one. Second, an SCX reaction is a necessary condition to end up with a
leading proton in the final state. 
We find that those SCX probabilities are very small (Sec.~\ref{sec:results}).

\subsection{$A(e,e'pN)$ cross-section ratios}
\label{sec:ratios}
Most often, it is extremely challenging to measure $A(e,e'pN)$ cross sections. 
A lot of information regarding nuclear SRC has been obtained by measuring cross-section rations
over extended ranges of the phase space \cite{Colle:massdep,Egiyan:2003vg,Egiyan:2005hs,Hen:2012jn,Hen31102014}.
Using the factorized form of the differential cross section from
Eq.~(\ref{eq:eeNNfactorized}), one can investigate $A(e,e'pN)$
cross-section ratios relative to $^{12}$C. Let $\widetilde{R}^{pN}_{A}$ ($R^{pN}_{A}$) be the
cross-section ratios with (without) the inclusion of FSI, 
\begin{align}
\widetilde{R}^{pN}_{A} = \frac{\widetilde{\sigma}_{A}^{pN}}{\widetilde{\sigma}_{^{12}\text{C}}^{pN}} \approx 
\frac{\int \textrm{d}^{2} \Omega_{k_{e'}} \textrm{d}^{3}\vec{k}_{12}
\sigma_{epN}(\vec{k}_{12}) \int \textrm{d}^{3}
\vec{P}_{12} F_{A}^{pN,D}(\vec{P}_{12})}{\int \textrm{d}^{2}
\Omega_{k_{e'}} \textrm{d}^{3}\vec{k}_{12}
\sigma_{epN}(\vec{k}_{12})\int \textrm{d}^{3} \vec{P}_{12}
F^{pN,D}_{^{12} \textrm{C}}(\vec{P}_{12})} 
= \frac{\int \textrm{d}^{3} \vec{P}_{12} F_{A}^{pN,D}(\vec{P}_{12})}{\int
\textrm{d}^{3} \vec{P}_{12} F^{pN,D}_{^{12} \textrm{C}}(\vec{P}_{12})} \, .
\label{eq:pN_ratios}
\end{align}
The $\widetilde{\sigma}_{A}^{pN}$ denotes the FSI-corrected $A(e,e'pN)$ cross section.
The cross-section ratios are independent of the information contained
in the photon-nucleon coupling
$\sigma_{epN}(\vec{k}_{12})$. Therefore, we use
cross-section ratios to quantify the effect of SRC as those are less
model dependent. 

In the limit of vanishing FSI the integrated c.m.~momentum
distribution $\int \textrm{d}^{3} \vec{P}_{12} \,
F^{pN,D}_{A}(\vec{P}_{12})$ is proportional to the amount of
SRC-susceptible $pN$ pairs. The relative amount of SRC pairs for nucleus
$A$ relative to $^{12}$C is then given by $ R_{pN} = \sigma_{A}^{pN} /
\sigma_{^{12}\textrm{C}}^{pN}$, where $\sigma_{A}^{pN}$ denotes the
$A(e,e'pN)$ cross section in the limit of vanishing FSI.
It is well established that the tensor correlation
\cite{Vanhalst:2014cqa, Subedi:2008zz} induces a heavy dominance of
SRC $pn$ pairs over SRC $pp$ pairs. This dominance is not
automatically generated in the ZRA without introducing additional
assumptions with regard to the dynamical mechanisms underlying the
SRC. The $pn$- over $pp$-pair dominance can be included for nucleus
$A$ using the measured $pn$/$pp$ pair ratio $(18 \pm 5)$ in $^{12}$C 
\cite{Subedi:2008zz}, in the following way
\begin{align}
	\frac{\sigma_{A}^{pn}}{\sigma_{A}^{pp}} =
        \frac{\sigma_{A}^{pn}}{\sigma_{^{12}\textrm{C}}^{pn}}
        \frac{\sigma_{^{12}\textrm{C}}^{pn}}{\sigma_{^{12}\textrm{C}}^{pp}}
        \frac{\sigma_{^{12}\textrm{C}}^{pp}}{\sigma_{A}^{pp}} \approx
        \frac{\sigma_{A}^{pn}}{\sigma_{^{12}\textrm{C}}^{pn}}
        \frac{\#pn\textrm{-pairs} \left( ^{12}\textrm{C} \right) }{2
          \cdot \#pp\textrm{-pairs} \left( ^{12}\textrm{C} \right)}
        \frac{\sigma_{^{12}\textrm{C}}^{pp}}{\sigma_{A}^{pp}} \approx
        \frac{\sigma_{A}^{pn}}{\sigma_{^{12}\textrm{C}}^{pn}} \left(
        \frac{18 \pm 5}{2} \right)
        \frac{\sigma_{^{12}\textrm{C}}^{pp}}{\sigma_{A}^{pp}} \, .
        \label{eq:pn_pp_dominance}
\end{align}

The exchanged photon can couple to both protons in a $pp$ pair and
to one in a $pn$ pair leading to the factor 2 in the denominator of Eq.~(\ref{eq:pn_pp_dominance}).
The expressions for the FSI-corrected cross-section ratios, $\widetilde{R}^{pN}=
\widetilde{\sigma}^{pN}_{A}/
\widetilde{\sigma}^{pN}_{^{12}\textrm{C}}$, are then given by (see
Fig.~\ref{fig:SCX_RMSGA_diag})
\begin{multline}
	\widetilde{R}^{pp}_{A} =
        \frac{\widetilde{\sigma}^{pp}_{A}}{\widetilde{\sigma}^{pp}_{^{12}\textrm{C}}}
        = \frac{ P_{\textrm{CX,A}}^{12,pp} T_{A}^{pp} \sigma^{pp}_{A}
          + P_{\textrm{CX,A}}^{1 \mathbf{[2]},pn} T^{p*}_{A}
          \sigma^{pn}_{A} }{ P_{\textrm{CX}^{12}\textrm{C}}^{12,pp}
          T_{^{12}\textrm{C}}^{pp} \sigma^{pp}_{^{12}\textrm{C}} +
          P_{\textrm{CX}^{12}\textrm{C}}^{1 \mathbf{[2]},pn}
          T^{p*}_{^{12}\textrm{C}} \sigma^{pn}_{^{12}\textrm{C}} }
         = \frac{ P_{\textrm{CX,A}}^{12,pp} T_{A}^{pp} R^{pp}_{A} +
          P_{\textrm{CX,A}}^{1 \mathbf{[2]},pn} T^{p*}_{A}
          \frac{\sigma^{pn}_{A}}{\sigma^{pp}_{^{12}\textrm{C}} } }{
          P_{\textrm{CX}^{12}\textrm{C}}^{12,pp}
          T_{^{12}\textrm{C}}^{pp} + P_{\textrm{CX}^{12}\textrm{C}}^{1
            \mathbf{[2]},pn} T^{p*}_{^{12}\textrm{C}}
          \frac{\sigma^{pn}_{^{12}\textrm{C}}}{\sigma^{pp}_{^{12}\textrm{C}}
        }} \\ = \frac{ P_{\textrm{CX,A}}^{12,pp} T_{A}^{pp} R^{pp}_{A}
          + P_{\textrm{CX,A}}^{1 \mathbf{[2]},pn} T^{p*}_{A}
          R^{pn}_{A} \frac{18 \pm 5}{2} }{
          P_{\textrm{CX}^{12}\textrm{C}}^{12,pp}
          T_{^{12}\textrm{C}}^{pp} + P_{\textrm{CX}^{12}\textrm{C}}^{1
            \mathbf{[2]},pn} T^{p*}_{^{12}\textrm{C}} \frac{18 \pm
            5}{2}}\, .
\label{eq:R_pp_FSI}
\end{multline}
Here, $\widetilde{\sigma}^{pN}_{A}$ is the FSI corrected $A(e,e'pN)$
cross section.  The first term in the numerator and denominator
consists of the $A(e,e'pp)$ cross section ($\sigma^{pp}_{A}$)
corrected for attenuation ($T^{pp}_{A}$) given that no SCX occurred
($P_{\textrm{CX,A}}^{12,pp}$). The second term is the contribution from an
initial $A(e,e'pn)$ ($\sigma^{pn}_{A}$) multiplied by the attenuation
factor ($T^{p*}_{A}$) given that the recoiling partner changes to a
proton ($P_{\textrm{CX,A}}^{1 \mathbf{[2]},pn}$). These two terms
correspond with the two possible paths to
$\widetilde{\sigma}^{pp}_{A}$ in Fig.~\ref{fig:SCX_RMSGA_diag} :
\begin{equation}
\sigma^{pN}_{A} \rightarrow \sigma^{pp}_{A}
\xrightarrow[]{P_{\textrm{CX,A}}^{12,pp} T^{pp}_{A}}
\widetilde{\sigma}^{pp}_{A} \; \; \; \text{and,} \; \; \; \sigma^{pN}_{A} \rightarrow
\sigma^{pn}_{A} \xrightarrow[]{P_{\textrm{CX,A}}^{1 \mathbf{[2]},pn}
  T^{p*}_{A}} \widetilde{\sigma}^{pp}_{A} \; .  
\end{equation}
In the ZRA, the nuclear $A(e,e'pN)$ transparency
$T^{pN}_{A}$ can be calculated as,
\begin{align}
T^{pN}_{A} \approx \frac{\int \textrm{d}^3 \vec{P}_{12} \,
  F^{pN,D}_{A}(\vec{P}_{12})}{\int \textrm{d}^3 \vec{P}_{12} \,
  F^{pN}_{A}(\vec{P}_{12})} \, .
\label{eq:transparency}
\end{align}
Here, $F^{pN}_{A}(\vec{P}_{12})$ is the c.m.~momentum distribution in the
limit of vanishing attenuation ($\mathcal{F}_{\text{RMSGA}} \equiv 1$
in Eq.~(\ref{eq:FD_RMSGA})). We stress that the transparency depends
on the sampled phase space, i.e. the integration volume of
$\vec{P}_{12}$ in Eq.~(\ref{eq:transparency}).

In estimating the attenuation effect for the SCX contribution we use
the averaged transparency $T^{p*}_{A} =
\frac{1}{2}(T^{pp}_{A}+T^{pn}_{A})$. The reason is that in our model
we have no information about the time ordering of the SCX and
attenuation mechanisms.  Therefore, starting from initial $pp$
knockout followed by $p \rightarrow n$ SCX, one can adopt the
averaged attenuation $\frac{T^{pn}_{A} + T^{pp}_{A}}{2}$.  Note that
the difference between $T^{pp}_{A}$ and $T^{pn}_{A}$ is rather small
for the kinematics addressed in this paper ($2 \%$ for
$^{12}\textrm{C}$ and about $20 \%$ for $^{208}\textrm{Pb}$).

For the $A(e,e'pn)$ cross-section ratios we get,
\begin{align}
\widetilde{R}^{pn}_{A} =
\frac{\widetilde{\sigma}^{pn}_{A}}{\widetilde{\sigma}^{pn}_{^{12}\textrm{C}}}
= \frac{ P_{\textrm{CX,A}}^{12,pn} T_{A}^{pn} \sigma^{pn}_{A} +
  P_{\textrm{CX,A}}^{1 \mathbf{[2]},pp} T^{p*}_{A} \sigma^{pp}_{A} }{
  P_{\textrm{CX}^{12}\textrm{C}}^{12,pn} T_{^{12}\textrm{C}}^{pn}
  \sigma^{pn}_{^{12}\textrm{C}} + P_{\textrm{CX}^{12}\textrm{C}}^{1
    \mathbf{[2]},pp} T^{p*}_{^{12}\textrm{C}}
  \sigma^{pp}_{^{12}\textrm{C}} }  = \frac{
  P_{\textrm{CX,A}}^{12,pn} T_{A}^{pn} R^{pn}_{A} +
  P_{\textrm{CX,A}}^{1 \mathbf{[2]},pp} T^{p*}_{A} R^{pp}_{A}
  \frac{2}{18 \pm 5} }{ P_{\textrm{CX}^{12}\textrm{C}}^{12,pn}
  T_{^{12}\textrm{C}}^{pn} + P_{\textrm{CX}^{12}\textrm{C}}^{1
    \mathbf{[2]},pp} T^{p*}_{^{12}\textrm{C}} \frac{2}{18 \pm 5} }\, .
\label{eq:R_pn_FSI}
\end{align}
As in Eq.~(\ref{eq:R_pp_FSI}) each term can be identified with a
certain path to $\widetilde{\sigma}^{pn}_{A}$ in
Fig.~\ref{fig:SCX_RMSGA_diag}.  The experimental values for
$\widetilde{R}^{pn}_{A}$ are not known if the outgoing neutrons are
not detected. In kinematics tuned so that the $A(e,e'p)$ signal is
dominated by $A(e,e'pN)$ events it is possible to deduce
$\widetilde{R}^{pn}_{A}$ from the $A(e,e'p)$ cross section ratios
($\widetilde{R}^{p}_{A}$) measured for the same kinematical settings
\begin{align}
	\widetilde{R}^{p}_{A} =
        \frac{\widetilde{\sigma}^{p}_{A}}{\widetilde{\sigma}^{p}_{^{12}\textrm{C}}}
        \approx \frac{ 2 \widetilde{\sigma}^{pp}_{A} +
          \widetilde{\sigma}^{pn}_{A}}{2\widetilde{\sigma}^{pp}_{^{12}\textrm{C}}
          + \widetilde{\sigma}^{pn}_{^{12}\textrm{C}}} = \frac{ 2
          \widetilde{R}^{pp}_{A} + \widetilde{R}^{pn}_{A}
          \widetilde{R}^{\frac{pn}{pp}}_{^{12}\textrm{C}}}{2 +
          \widetilde{R}^{\frac{pn}{pp}}_{^{12}\textrm{C}}} \, .
          \label{eq:R_p}
\end{align}

The $^{12}$C$(e,e'pn)$ over $^{12}$C$(e,e'pp)$ cross-section ratio 
$ \widetilde{R}^{\frac{pn}{pp}}_{^{12}\textrm{C}}$ can be extracted in the following way
\begin{align}
	\widetilde{R}^{\frac{pn}{pp}}_{^{12}\textrm{C}} =
        \frac{\widetilde{\sigma}^{pn}_{^{12}\textrm{C}}}{\widetilde{\sigma}^{pp}_{^{12}\textrm{C}}}
        = \frac{ P_{\textrm{CX}^{12}\textrm{C}}^{12,pn}
          T_{^{12}\textrm{C}}^{pn} \sigma^{pn}_{^{12}\textrm{C}} +
          P_{\textrm{CX}^{12}\textrm{C}}^{1 \mathbf{[2]},pp}
          T^{p*}_{^{12}\textrm{C}} \sigma^{pp}_{^{12}\textrm{C}} }{
          P_{\textrm{CX}^{12}\textrm{C}}^{12,pp}
          T_{^{12}\textrm{C}}^{pp} \sigma^{pp}_{^{12}\textrm{C}} +
          P_{\textrm{CX}^{12}\textrm{C}}^{1 \mathbf{[2]},pn}
          T^{p*}_{^{12}\textrm{C}} \sigma^{pn}_{^{12}\textrm{C}} }
          = \frac{ P_{\textrm{CX}^{12}\textrm{C}}^{12,pn}
          T_{^{12}\textrm{C}}^{pn} \frac{18 \pm 5}{2} +
          P_{\textrm{CX}^{12}\textrm{C}}^{1 \mathbf{[2]},pp}
          T^{p*}_{^{12}\textrm{C}}}{
          P_{\textrm{CX}^{12}\textrm{C}}^{12,pp}
          T_{^{12}\textrm{C}}^{pp} + P_{\textrm{CX}^{12}\textrm{C}}^{1
            \mathbf{[2]},pn} T^{p*}_{^{12}\textrm{C}} \frac{18 \pm
            5}{2} } \, .
\end{align}
Hence from Eq.~(\ref{eq:R_p}),
\begin{align}
	\widetilde{R}^{pn}_{A} =
        \frac{1}{\widetilde{R}^{\frac{pn}{pp}}_{^{12}\textrm{C}}}
        \left[ \widetilde{R}^{p}_{A} \left( 2 +
          \widetilde{R}^{\frac{pn}{pp}}_{^{12}\textrm{C}} \right) - 2
          \widetilde{R}^{pp}_{A} \right] \, .
\end{align}
The relations for $\widetilde{R}^{pp}_{A}$ and
$\widetilde{R}^{pn}_{A}$ can be inverted to extract the
FSI-uncorrected cross section ratios (which are proportional to the
ratios of SRC prone pairs) $R^{pp}_{A}$, $R^{pn}_{A}$ from the
measured values for $\widetilde{R}^{pp}$ and
$\widetilde{R}^{pn}$~\cite{Colle:massdep}.

\section{Results}
\label{sec:results}
In this section we present the results of the numerical $A(e,e'pN)$
calculations for four representative target nuclei and two
representative but distinct kinematic settings.  First, we apply the
formalism developed in the previous section to the $A(e,e'pN)$
reaction in the kinematics covered by the Jefferson Lab CLAS detector
\cite{Hen31102014}.  The latter is a ``$4\pi$'' detector, which
results in a very large phase-space coverage. We systematically refer
to this kinematics as ``KinB''. Kinematics approaching a ``$4 \pi$''
layout pose challenges for the calculations and require dedicated
sampling techniques that are outlined below. After the discussion of
the ``$4\pi$'' KinB results we present two-nucleon knockout
calculations in kinematics in very narrow
solid angles for all detected particles (coined ``KinA'').

In dealing with the KinB situation, we define a reference frame with
the $z$-axis along the initial momentum $\vec{k}_1$ of the proton and
the exchanged photon-momentum $\vec{q}$ in the $x-z$ plane.  A
two-nucleon knockout event is uniquely characterized by the set of 6
kinematical variables $\{Q^{2} \equiv |\vec{q} \,|^2 - \omega ^2, x_B = \frac
{Q^2} {2 m_N \omega}, \theta_{q},\vec{P}_{12}\}$. Here, $\theta_{q}$
is the direction of $\vec{q}$ relative to the $z$ axis.

Upon numerically computing the distorted c.m. momentum distribution of
Eq.~(\ref{eq:FD_RMSGA}), we generate phase space samples by drawing
($x_B$,$Q^{2}$) from the experimentally measured $(x_B,Q^2)$
distribution~\cite{Hen:2012jn}.  We draw the $\theta_{q}$ and the
$\vec{P}_{12}$ uniformly from the relevant ranges.  In order to
guarantee that the virtual photon primarily probes correlated pairs a
number of kinematic constraints are imposed
\begin{align}
	& \theta_{\vec{p}_1,\vec{q}} \leq 25^{\circ} \; ; \; 0.62 <
  \frac{|\vec{p}_1|}{|\vec{q}|} < 0.96 \; ; \; x_B \geq 1.2 ; \\ &
  |\vec{k}_1| \geq 300 \textrm{ MeV} ; |\vec{p}_2| \geq 350 \textrm{
    MeV} \, .
	\label{eq:kinematic_cuts}
\end{align}
The first two cuts select events where the virtual photon has mainly
interacted with the struck (leading) proton.  The $x_{B} \geq 1.2$ cut
selects events with a high $|\vec{q}|$ and relatively low $\omega$,
suppressing for example pion production through intermediate $\Delta$
production.  The last two cuts impose high-momentum conditions (larger
than the Fermi momentum) for the initial nucleon pair.

\begin{figure} [ht]
\centering
\includegraphics[width=0.7\columnwidth]{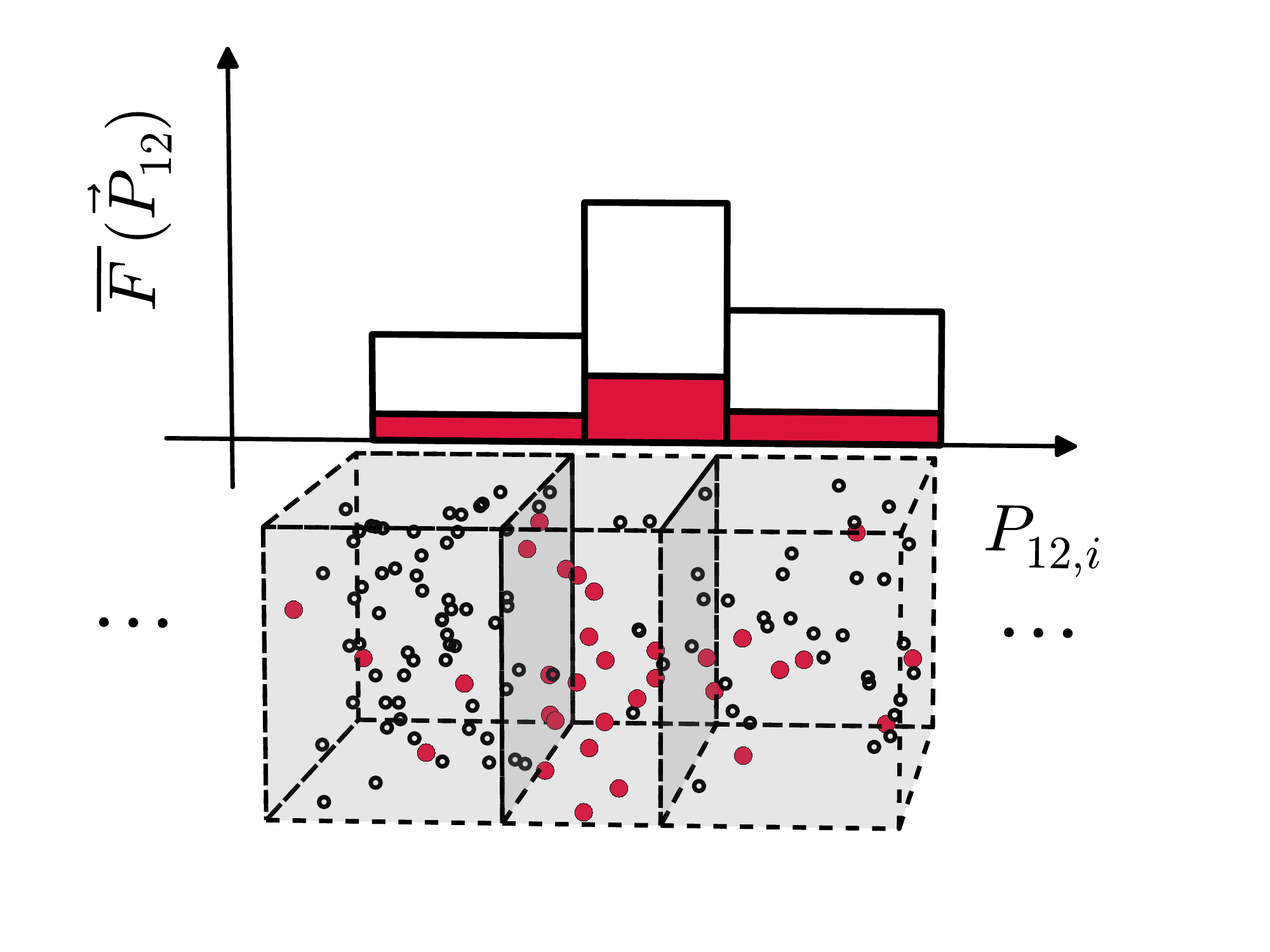}
\includegraphics[width=0.6\columnwidth]{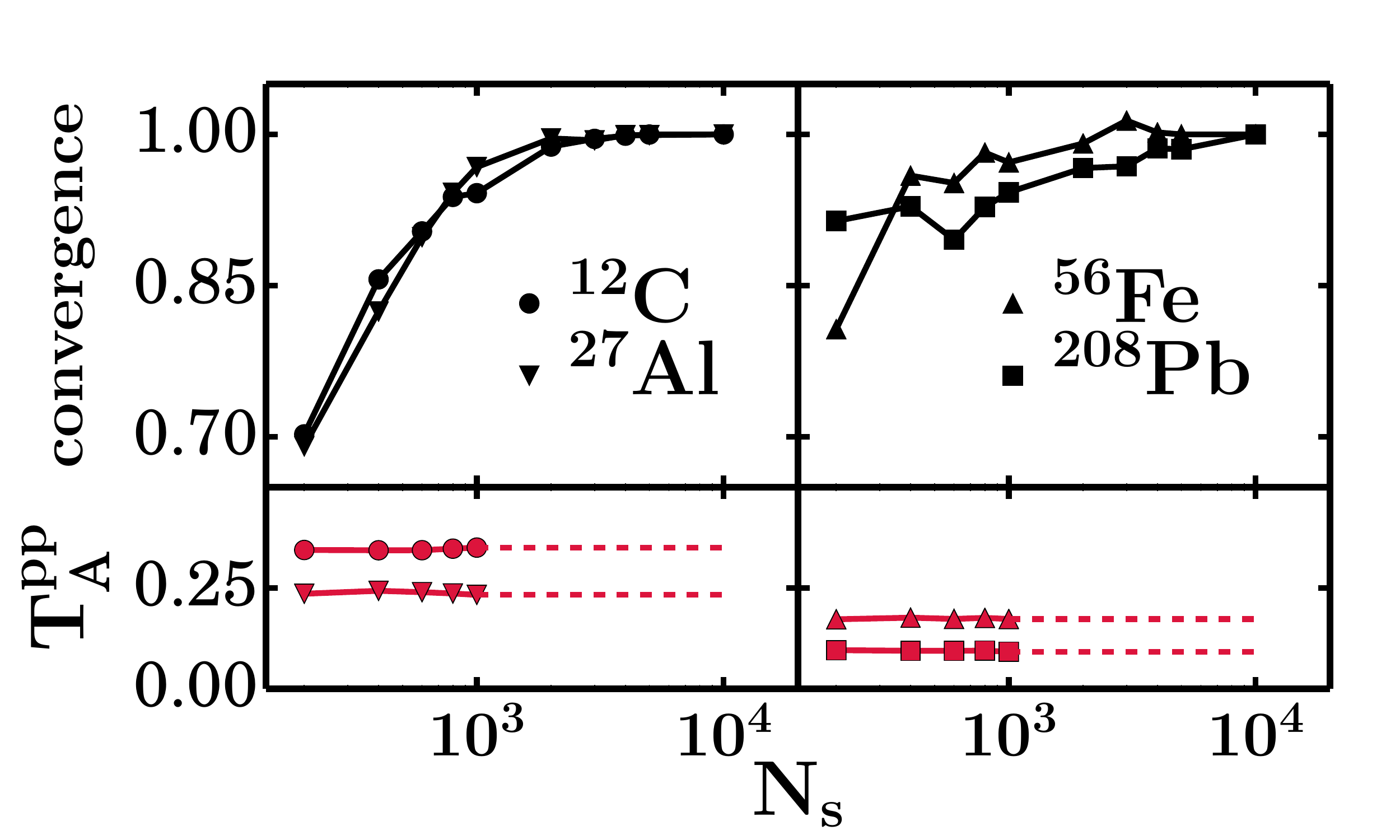}
\caption{(Color online) (Top) A schematic representation of the
  sampling procedure adopted in the $A(e,e'pN)$ calculations in
  kinematics covering a large phase space. The gray cubes are
  $\vec{P}_{12}$ bins.  The black circles are those points in
  $\vec{P}_{12}$ space for which $A(e,e'pN)$ calculations with
  vanishing FSI ($\mathcal{F}_{\textrm{RMSGA}} \equiv 1$) are
  done. The red dots represent the sampled points for which the RMSGA
  $A(e,e'pN)$ calculations are performed. The sampling weight of each
  bin is the bin-averaged c.m.~momentum distribution
  $\overline{F}^{pN}_{A}(\vec{P}_{12})$ indicated by the white
  bars. The resulting bin-averaged c.m.~momentum distribution
  including FSI is indicated by the red bars.
(Bottom) The convergence as defined in
  Eq.~(\ref{eq:convergence_constant_def}) and the transparency
  $T^{pp}_{A}$ of Eq.~(\ref{eq:transparency}) as a function of the
  sample size $N_{s}$ for $(e,e'pp)$ from $^{12}$C, $^{27}$Al,
  $^{56}$Fe and $^{208}$Pb in the kinematics defined by
  Eq.~(\ref{eq:kinematic_cuts}).}
\label{fig:sampling_pict_convergence}
\end{figure}

Sampling over the complete $\{x_B$,$Q^2$,$\theta_{q}$,$\vec{P}_{12}\}$
space is computationally very demanding in the RMSGA
calculations. Therefore, we use stratified sampling on the binned
plane-wave result $F^{pN}_{A}(\vec{P}_{12}) = F^{pN,D}_{A}(\vec{P}_{12} |
\mathcal{F}_{\mathrm{RMSGA}} \equiv 1 ) $, to generate the events in
phase space (see Fig.~\ref{fig:sampling_pict_convergence} for an
illustration).  Thereby, after calculating $F^{pN}_{A}(\vec{P}_{12})$ for a
large number of events, we bin the events in the 
$\vec{P}_{12}$ space.  Next, we sample phase-space events from the
bins using the bin-averaged value
$\overline{F}^{pN}_{A}(\vec{P}_{12})$ of $F^{pN}_{A}(\vec{P}_{12})$ as
bin weights.  We then include the effect of attenuation by
calculating $F^{pN,D}_{A}(\vec{P}_{12})$ in the RMSGA, for the
sampled phase-space events. It is assumed that the bin averaged
$\overline{F}^{pN,D}_{A}(\vec{P}_{12})$ of the function
$F^{pN,D}_{A}(\vec{P}_{12})$ of the sampled events is representative
for the real bin average. Using this procedure, the integrals in
Eq.~(\ref{eq:pN_ratios}) are determined in the following way:
\begin{align}
	\int \textrm{d}^3 \vec{P}_{12} F^{pN,D}_{A}(\vec{P}_{12})
        \approx \frac{V_{\vec{P}_{12}}}{N_s} \sum_{n \in
          \textrm{bins}} \overline{F}^{pN,D}_{A,n}(\vec{P}_{12})
        N_{n}\,,
\end{align}
with $N_n$ the number of events in the $n$-th bin, $N_{s}$ the total
number of phase-space events and $V_{\vec{P}_{12}}$ the considered
phase-space volume in $\vec{P}_{12}$.

Fig.~\ref{fig:sampling_pict_convergence} displays the convergence of
the plane-wave integrated c.m.~distribution, defined as,
\begin{align}
	\frac{\left[ \int \textrm{d}^{3} \vec{P}_{12}
            F^{pN}_{A}(\vec{P}_{12}) \right]_{N_s}}{\left[ \int
            \textrm{d}^{3} \vec{P}_{12} F^{pN}_{A}(\vec{P}_{12})
            \right]_{N_s=10^{4}} } \, ,
	\label{eq:convergence_constant_def}
\end{align}
and the nuclear transparency $T^{pp}_{A}$
(Eq.~(\ref{eq:transparency})) as a function of the number of sampled
events $N_s$.  The convergence at a 1000 samples is between 94\% and
97\% for all nuclei. We perform the RMSGA calculations for this sample
size. From Fig.~\ref{fig:sampling_pict_convergence} it is clear that
the nuclear transparency is almost independent of the sample
size. This indicates that the ZRA and RMSGA ($\equiv$ZRA+RMSGA) have
almost identical convergence behavior as a function of the sample size $N_s$.

\begin{figure*}
\centering
	\includegraphics[width=0.8\textwidth]{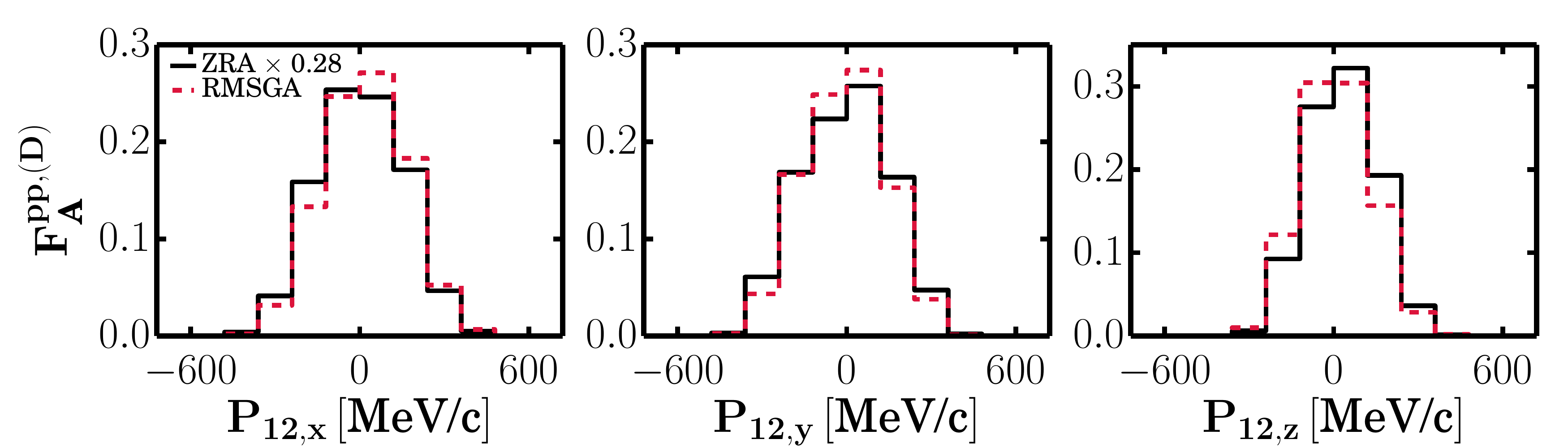}
	\includegraphics[width=0.8\textwidth]{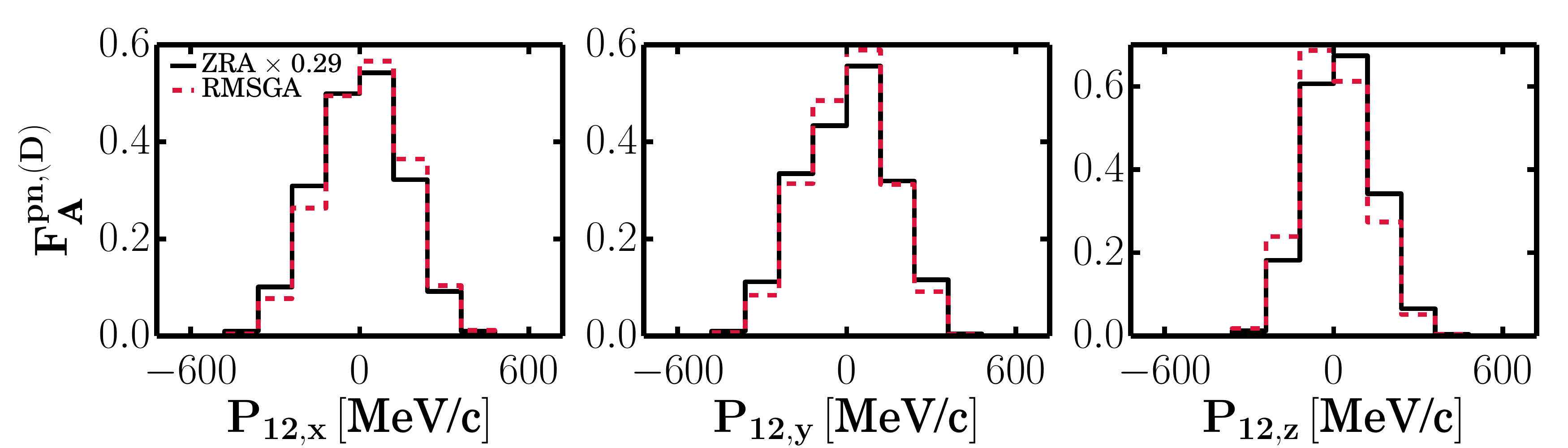}
	\caption{(Color online) The $P_{12,x}$,$P_{12,y}$,$P_{12,z}$
          dependence of the c.m.~momentum distribution
          $F^{pN}_{A}(\vec{P}_{12})$ (ZRA in the plane-wave limit of
          the ejected nucleons) and $F^{pN,D}_{A}(\vec{P}_{12})$
          (including elastic attenuation of the ejected nucleons) for
          $^{12}$C$(e,e'pp)$ (top) and $^{12}$C$(e,e'pn)$
          (bottom) in KinB kinematics. The solid lines, obtained in the ZRA are
          multiplied with the nuclear transparency for
          $^{12}\textrm{C}$ (Table~\ref{tab:ratios_transparencies}).}
	\label{fig:cm_distribution_comp}
\end{figure*}

\begin{figure*}
\centering
	\includegraphics[width=0.49\textwidth]{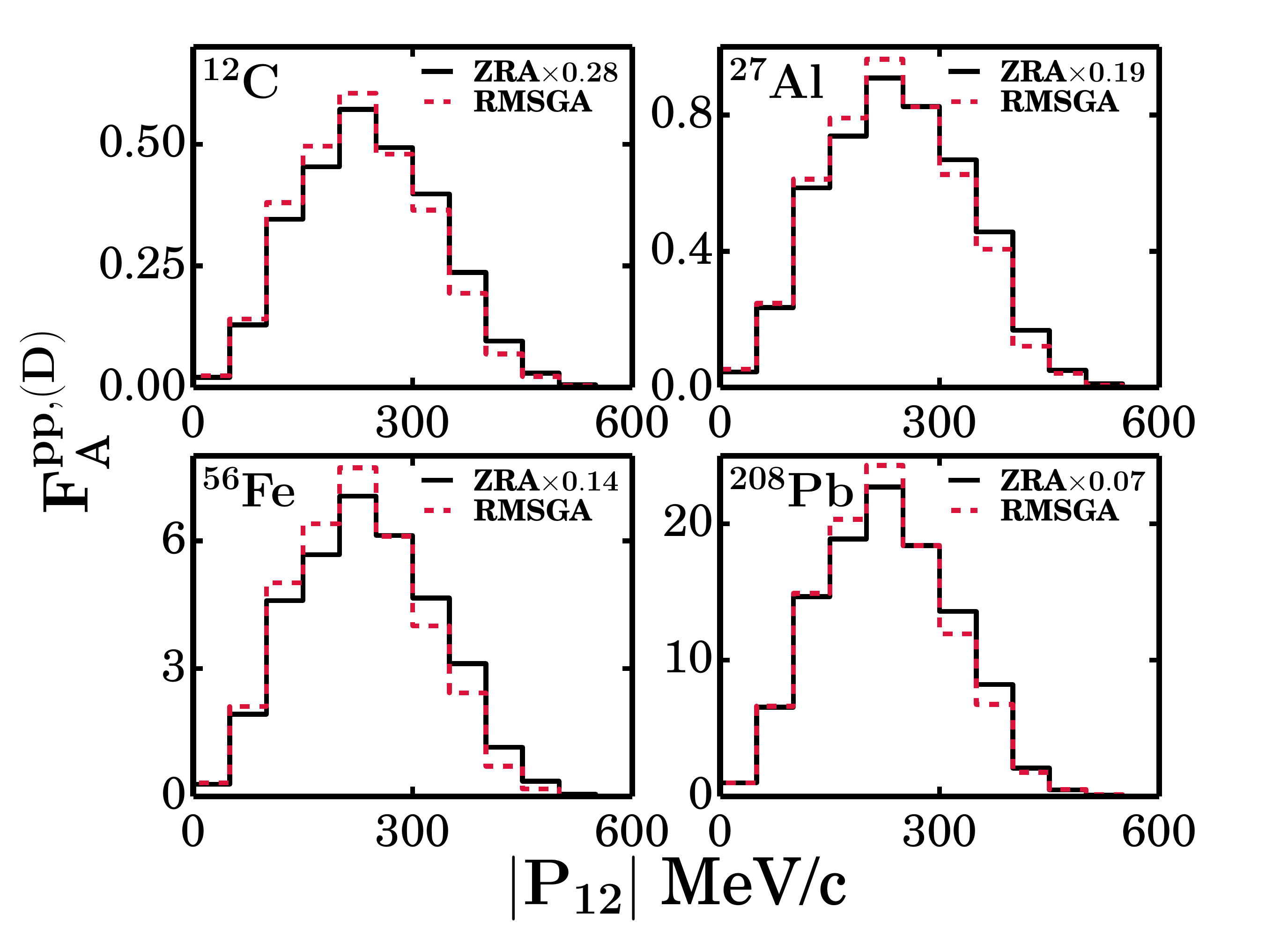}
	\includegraphics[width=0.49\textwidth]{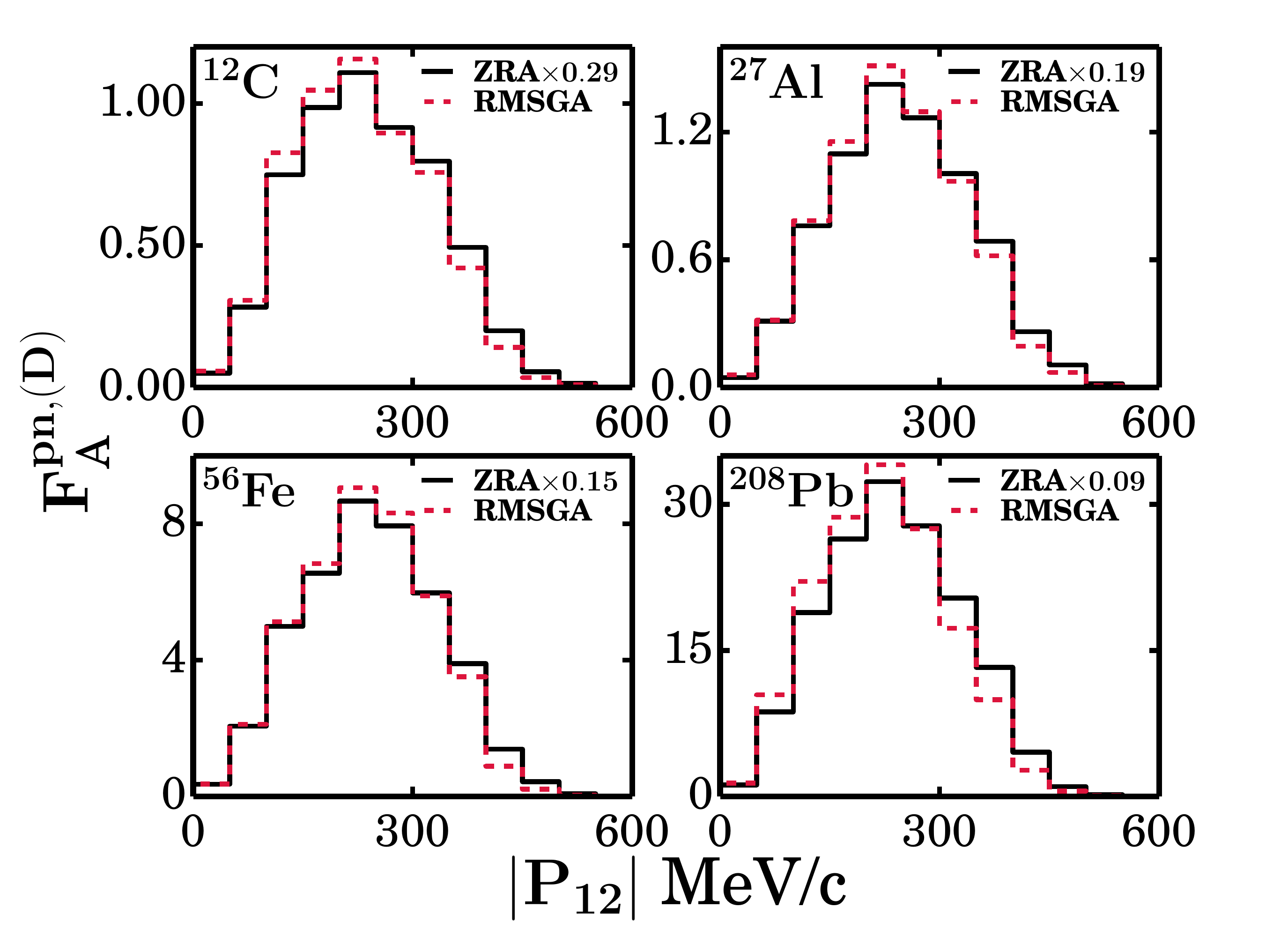}
	\caption{(Color online) The c.m.~momentum distribution in the
          ZRA $F^{pN,(D)}_{A}(\vec{P}_{12})$ with and without RMSGA attenuation 
          corrections for $A(e,e'pp)$ (left) and $A(e,e'pn)$ (right) in KinB kinematics. As
          in Fig.~\ref{fig:cm_distribution_comp} the ZRA results are
          multiplied with the corresponding $T_{A}^{pN}$ (see Table
          \ref{tab:ratios_transparencies}).}
	\label{fig:cm_distribution_mag}
\end{figure*}

%
%
Figures~\ref{fig:cm_distribution_comp}
and~\ref{fig:cm_distribution_mag} show the computed c.m.~distribution
for $A(e,e'pp)$ and $A(e,e'pn)$.  Both undistorted (ZRA,
$F^{pN}(\vec{P}_{12})$) and distorted (RMSGA,
$F^{pN,D}(\vec{P}_{12})$) results are shown. It is clear that for all
target nuclei considered, attenuation effects on the ejected nucleons
marginally affect the shape of the c.m.~momentum distribution. Note
that the shape of the c.m.~momentum distribution is fairly similar for
all four nuclei considered. This illustrates that SRC are connected
with the local and ``universal'' short-distance behavior of nucleon
pairs \cite{Vanhalst:2014cqa}.

\begin{figure}
\centering
\includegraphics[width=0.8\textwidth]{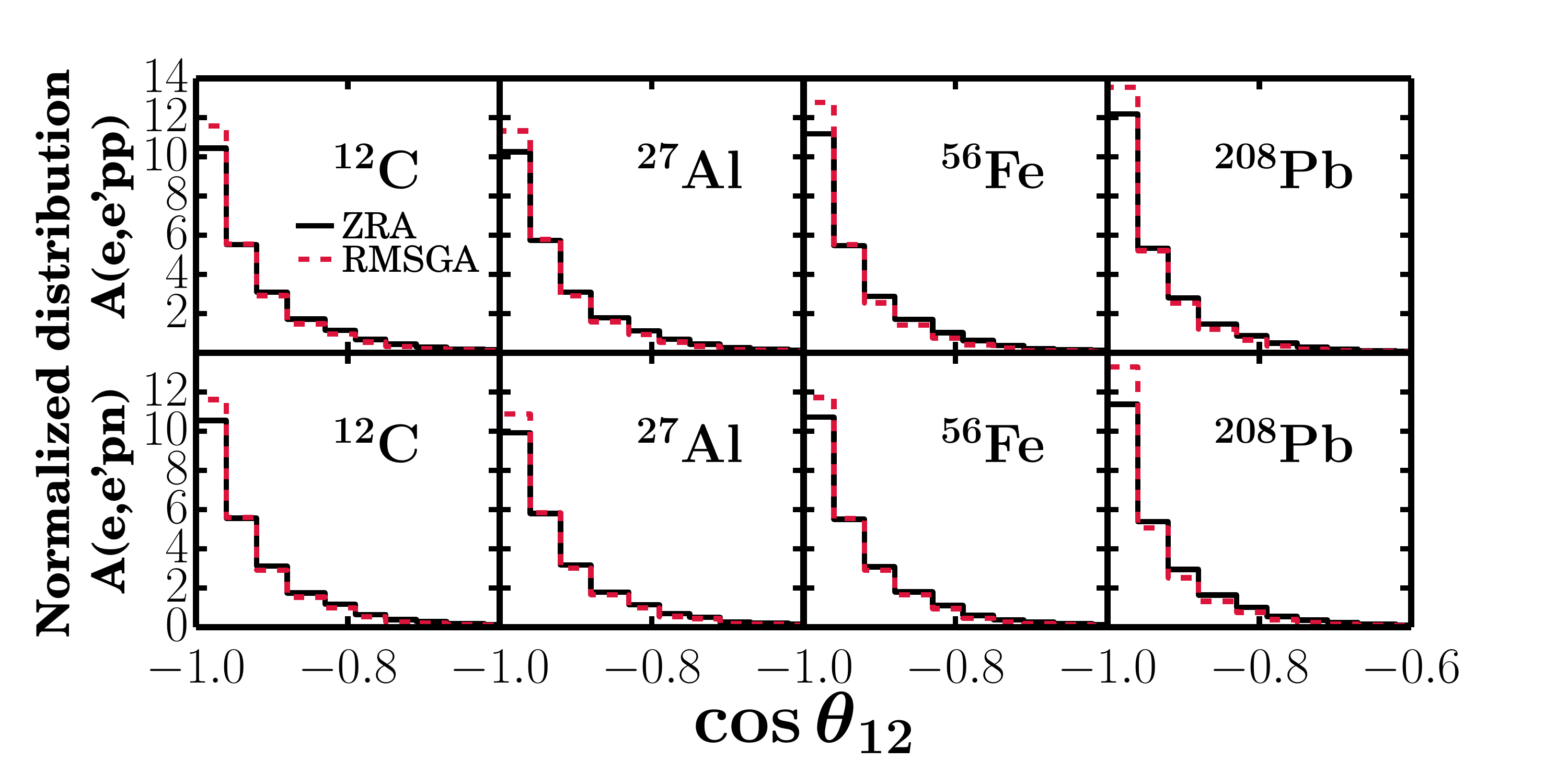}
\caption{(Color online) The normalized $\theta_{12}$ distributions for
  $A(e,e'pp)$ (top) and $A(e,e'pn)$ (bottom) in KinB kinematics.}
\label{fig:opening_angle}
\end{figure}

%
%
The opening angle $\theta _{12}$ is defined
as the angle between the two initial nucleon momenta, $\cos \theta
_{12} = ( \vec{k}_1 \cdot \vec{k}_2 ) / ( | \vec{k}_1 | | \vec{k}_2 |
)$. 
Nucleon pairs susceptible to SRC have a high relative momentum and a
small c.m momentum, reminiscent of ``back-to-back'' motion.  This
causes the opening-angle distribution of SRC pairs to be biased
towards backward angles. Figure~\ref{fig:opening_angle} displays the normalized $\theta_{12}$ 
distributions as they can be extracted from the undistorted and
distorted distributions $F^{pN}(\vec{P}_{12})$ and
$F^{pN,D}(\vec{P}_{12})$ for the different nuclei.  The inclusion of
elastic attenuation mechanisms as computed in the RMSGA framework, has
a relatively small effect on the opening-angle distribution. A slight
tendency to effectively increase the contributions of the $\cos \theta
_{12} \approx -1 $ events is observed.

\begin{figure}
\centering
\includegraphics[width=0.7\textwidth]{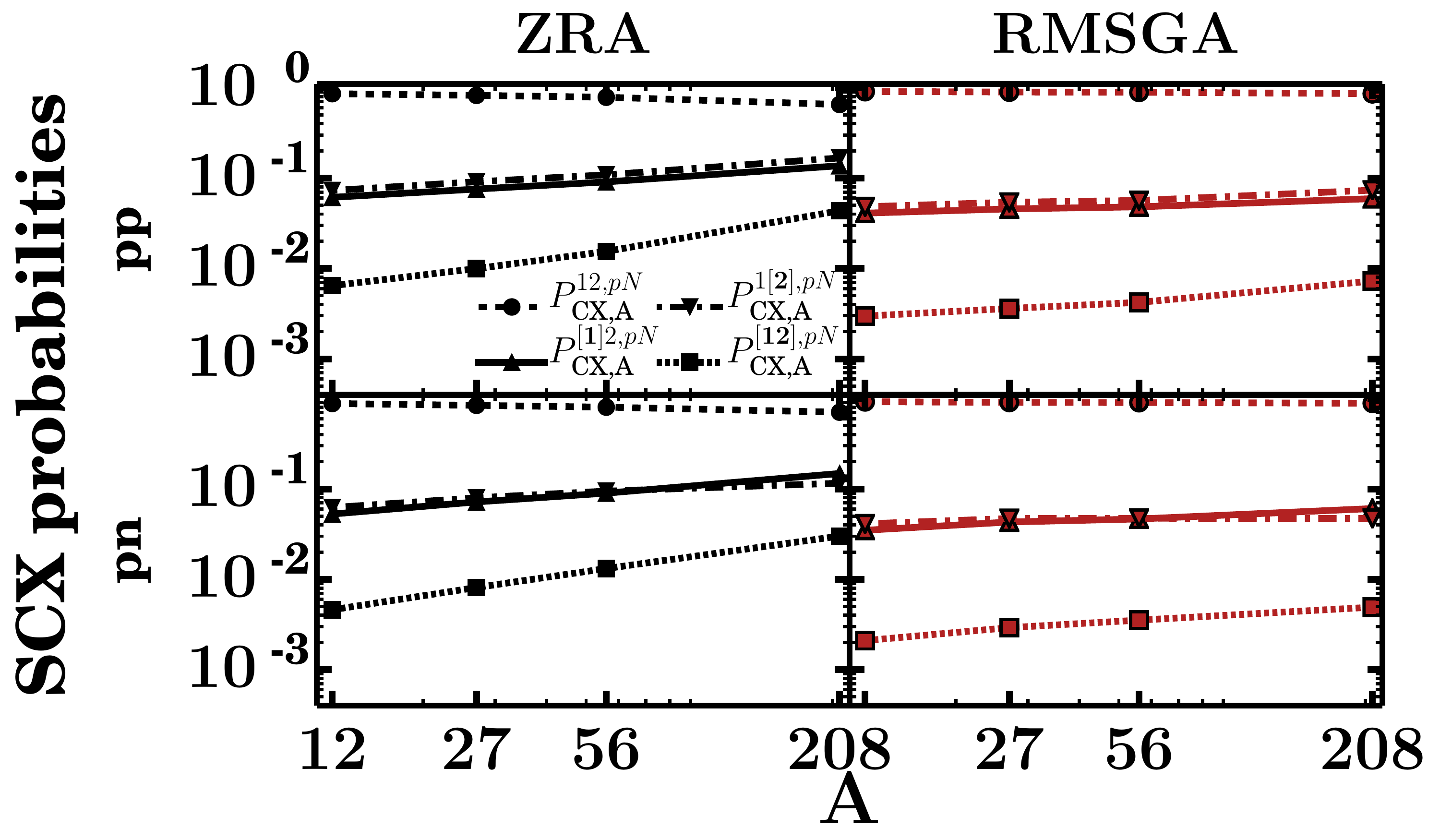}
\caption{(Color online) The mass dependence of the SCX probabilities for $A(e,e'pp)$
  (top) and $A(e,e'pn)$ (bottom) for KinB. $P_{\textrm{CX,A}}^{12,pN}$ is the
  probability that no charge exchange scattering
  occurs. $P_{\textrm{CX,A}}^{\mathbf{[1]},2,pN}$,
  $P_{\textrm{CX,A}}^{1,\mathbf{[2]},pN}$ are the probabilities that
  either the leading or the recoil nucleon undergoes charge
  exchange. $P_{\textrm{CX,A}}^{\mathbf{[1,2]},pN}$ is the probability
  that both the leading proton and recoiling nucleon undergo charge
  exchange. The black lines (left) are calculated in the ZRA, the red
  ones (right) include RMSGA attenuation.}
\label{fig:cx_probs}
\end{figure}

Charge-exchange reactions in the final state will mix the
c.m.~momentum distribution and the opening-angle distribution of the
initial $pp$- and $pn$-pairs.  For example initial $pp$-pairs, with a
c.m. momentum distribution $F^{pp,D}_{A}(\vec{P}_{12})$, can change
into $pn$-pairs, contaminating $F^{pn,D}_{A}(\vec{P}_{12})$ and vice
versa. From Figs.~\ref{fig:cm_distribution_comp} to
\ref{fig:opening_angle} it is clear that throughout the mass table 
the $A(e,e'pp)$ and $A(e,e'pn)$  c.m.~momentum distributions as well as the opening angle distributions
are very similar. The effect of SCX on the shape of these
distributions is close to negligible.
The SCX probabilities calculated in the ZRA and ZRA+RMSGA
(Eq.~\ref{eq:cx_prob_example}) are displayed in
Fig.~\ref{fig:cx_probs}.  The RMSGA clearly diminishes the SCX
probabilities. This can be understood in the following way:
the events most susceptible to SCX reactions are those whereby
the ejected nucleon pair traverses large distances in the recoiling
nucleus. These events are most suppressed by the attenuation, causing
the SCX probabilities to decrease.

Table~\ref{tab:ratios_transparencies} lists the cross-section ratios $R^{pN}_{A}$ calculated in the ZRA. These are approximately equal to the SRC pair ratios.
The nuclear transparencies $T^{pN}_{A}$ calculated in the RMSGA (Eq.~(\ref{eq:transparency})) and the FSI (RMSGA+SCX) corrected cross section ratios $\widetilde{R}^{pN}_{A}$ are listed as well. The RMSGA attenuates the cross sections
significantly, ranging from a factor of four for $^{12}$C to fourteen for $^{208}$Pb. The inclusion of SCX has a very modest effect on the cross section ratios $\widetilde{R}^{pN}_{A}$, the largest effect is approximately 8\% for $\widetilde{R}^{pp}_{^{208}\text{Pb}}$.

\begin{table}
\centering
\begin{tabular}{ c r r r r r r r } \hline \hline
 & $R^{pp}_{A}$ & $T_{A}^{pp}$ & \multicolumn{1}{c}{$\widetilde{R}^{pp}_{A}$} & $R^{pn}_{A}$ & $T_{A}^{pn}$ & \multicolumn{1}{c}{$\widetilde{R}^{pn}_{A}$} & $(T^{p}_{A})^2$\\ \hline
 $^{12}$C  &  1.00 & 0.280 & $1.00\phantom{^{+0.00}_{-0.00}}$  &  1.00 & 0.286 & $1.00\phantom{^{+0.00}_{-0.00}}$ & 0.26 \\
$^{27}$Al  &  2.89 & 0.186 & $1.91^{+0.01}_{-0.01}$ &  2.52 & 0.186 & $1.65^{+0.01}_{-0.01}$ & $-$  \\
$^{56}$Fe  &  5.89 & 0.138 & $2.85^{+0.01}_{-0.01}$ &  4.82 & 0.150 & $2.49^{+0.01}_{-0.01}$ & 0.10 \\
$^{208}$Pb & 17.44 & 0.073 & $4.96^{+0.11}_{-0.14}$ & 18.80 & 0.093 & $6.00^{+0.01}_{-0.01}$ & 0.05 \\ \hline \hline
\end{tabular}
\caption{The numerical results for the cross-section ratios
  $R^{pN}_{A}$ (ZRA) and the corresponding transparencies
  (Eq.~\ref{eq:transparency}) calculated with the
  RMSGA. $\widetilde{R}^{pN}_{A}$ are the cross section ratios corrected
  for FSI (RMSGA and SCX). For vanishing SCX probabilities $\widetilde{R}^{pN}_{A}$ is equal to $R^{pN}_{A}T_{A}^{pN}/T_{^{12}\textrm{C}}^{pN}$. 
  $T^{p}_{A}$ is the measured $A(e,e'p)$ transparency~\cite{Cosyn:2013qe}.}
\label{tab:ratios_transparencies}
\end{table}

\begin{figure}
\centering
\includegraphics[width=0.8\textwidth]{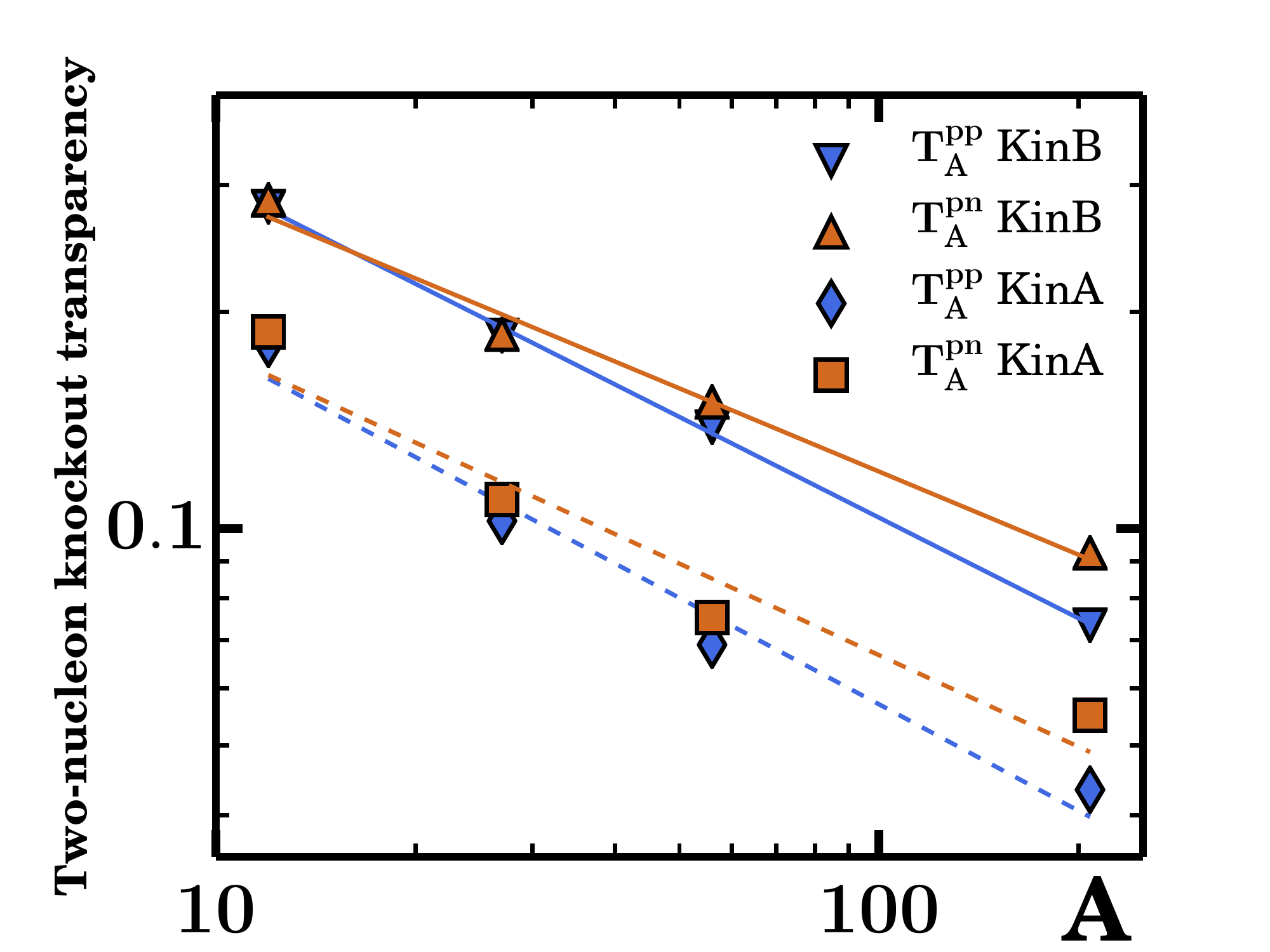}
\caption{(Color online) Mass dependence of the two-nucleon knockout transparency calculated according to Eq.
(\ref{eq:transparency}). The lines are power-law fits ($\sim A^{\lambda}$) to the numerical predictions. KinB denotes the results for the $A(e,e'pN)$ kinematics accessed in Ref.~\cite{Hen31102014} covering a large phase space. 
KinA are the transparencies calculated for $A(e,e'pN)$ in the kinematics accessed in
Ref.~\cite{Shneor:2007tu} with very selective phase space.}
\label{fig:transparencies}
\end{figure}

The mass dependence of the calculated transparencies for the ``$4
\pi$'' kinematics KinB follow a power law ($ T_{A}^{pN} \propto
A^{\lambda}$) and are displayed in Fig.~\ref{fig:transparencies}. Up to
now we concentrated on the kinematics accessed in
the experiment of Ref.~\cite{Hen31102014}.

We test the robustness of our methodology by applying it to the
kinematics accessed in the $^{12}$C$(e,e'pp)$ measurements of Ref.~\cite{Shneor:2007tu}, denoted KinA. It corresponds with
a very selective phase space whereby the scattered electron and leading proton are detected with two high resolution spectrometers at the fixed central angles 19.5$^{\circ}$ (electron) and -35.8$^{\circ}$ (leading proton) relative to the incoming electron beam. The angular acceptance is $\pm 0.03$ mrad ($\pm 0.06$ mrad) in the horizontal (vertical) plane. The initial (final) electron momentum is fixed at 4.627~GeV/c (3.724~GeV/c). The leading proton momentum is $1.42\pm4\%$ GeV/c. The recoiling proton is detected at the central angle -$99^{\circ}$ with an angular acceptance of 96 msr. These kinematics are finely tuned and optimized to select knockout reactions of initial back-to-back pairs. For example, more than 80\% of the available phase space has opening angle $\cos \theta_{12} < -0.9$.
 
The power-law dependencies of the $T_{A}^{pN}$
transparencies in KinA and KinB kinematics are very similar and are
included in Fig.~\ref{fig:transparencies}. We find $T^{pp}_{A}
\propto A^{-0.46\pm0.02}$ (KinB), $T^{pp}_{A} \propto
A^{-0.49\pm0.06}$ (KinA), $T^{pn}_{A} \propto A^{-0.38 \pm 0.03}$
(KinB) and $ T^{pn}_{A} \propto A^{-0.42 \pm 0.05} $ (KinA).  This
indicates that the mass dependence of the transparency is robust.
The absolute value of the KinA transparencies is lower by approximately a factor of 2 compared to the KinB results.
Given the small phase space of KinA we cannot make a detailed study
of the c.m.~momentum distribution $F^{pN,(D)}_{A}$ and the
opening-angle distribution, as was done for KinB. 
Indeed, KinA kinematics only covers restricted ranges in $\vec{P}_{12}$ and $\cos \theta_{12}$.

Next, we outline an alternative method to account for the mass dependence
of $T_{A}^{pN}$.  The transparency $T_{A}^{p}$ of $A(e,e'p)$ processes
can be interpreted as the probability of a single proton leaving the
nucleus after virtual photon excitation. Recent measurements
\cite{Hen:2012jn} have confirmed that the $A$ dependence of the
$T_{A}^{p}$ can be captured by the power-law $A^{ -0.33}$ \cite{Lava:2004} . One could
naively expect that $T_{A}^{pp} \approx T_{A}^{p} T_{A}^{p} \approx A
^{-0.66}$. Upon squaring the $T_{A}^{p}$ one assumes that the two
protons are independent. This is in obvious contradiction with the
ZRA picture for SRC-driven two-nucleon knockout reaction where the nucleon pair is maximally correlated: finding one
nucleon at the spatial coordinate $\vec{R}_{12}$ fixes the second
nucleon's spatial coordinate.
%
%
To obtain a deeper and more intuitive understanding of the $A$ dependence
of $T_{A}^{pN}$ we have developed a toy model detailed in
Appendix~\ref{app:toy_model}. Thereby the nucleus is treated as
a uniform sphere with radius $R = 1.2 A^{-\frac{1}{3}}$ fm and density
$\rho = \frac{A}{4/3 \pi R^{3}} = 0.138 \textrm{ fm}^{-3}$. We
calculate the transparencies using a semi-classical approach analogous
to the method used to compute the SCX probabilities. The attenuation
is derived using the scattering probabilities as in
Eq.~(\ref{eq:CX_prob_r}), where the scattering cross section is treated
as a model parameter. We derive a range $\lambda \in [-0.37,-0.78]$
for the exponent $\lambda$ in $T^{pN}_{A} \propto A^{\lambda}$.
This range is established by varying two
parameters: (1) the nucleon-nucleus cross section describing the
attenuation, (2) the $\theta_{12}$
distribution for the nucleon pair is varied from a uniform
distribution (no angular correlation) to a back-to-back delta function
$\delta(\theta_{12} - \pi)$.  The exponents derived
involving the full calculations (Fig.~\ref{fig:transparencies}) are
well in agreement with the toy model. 
The toy model predicts that for increasingly backward
peaked $\theta_{12}$ distributions the
exponent $\lambda$ becomes more negative.
The toy model also explains why the $T_{A}^{pN}$ diminishes as one increasingly selects back-to-back nucleons.

\section{Summary} \label{sec:conclusion}
We have studied the impact of final-state interactions in SRC-driven
exclusive $A(e,e'pN)$ processes.  Attenuation through elastic and soft
inelastic rescattering as well as single-charge exchange processes are
included in the description of the FSI. We
applied our model to two very different kinematics probing SRC pairs
and four target nuclei from carbon to lead. It is shown that the
inclusion of FSI has a limited effect on the extracted
shapes of the c.m.~momentum and opening-angle distributions of the
correlated nucleon pair. The cross section, however, is significantly
attenuated by the FSI. The absolute values of the transparencies
depend on the kinematics and we find $T^{pN}_{A} \approx 0.2 - 0.3 $
for a light nucleus like $^{12}$C and $T^{pN}_{A} \approx 0.04 - 0.07$
for a heavy nucleus like $^{208}$Pb.  The mass dependence of the
nuclear transparency is more robust. We find $T^{pp}_{A} \propto
A^{-0.46 \pm 0.02}$ and $T^{pn}_{A} \propto A^{-0.38 \pm 0.03}$ in
``$4 \pi$'' kinematics. For the highly selective kinematics, that
exclusively probes back-to-back nucleons, we find $T^{pp}_{A} \propto A^{-0.49 \pm 0.06}$
and $T^{pn}_{A} \propto A^{-0.42\pm0.05}$. Both are softer than one would expect from a
doubling of the power found for single-nucleon knockout ($ T_{\text{A}}^{p} \propto A^{-0.33}$). The values for the exponent $\lambda$ in the power-law dependence  of $T_{A}^{pN} \propto A^{\lambda}$
are tested against the results of a toy model which allows us to set bounds on values
for $\lambda$.  We find the calculated values to be well within these
bounds, $\lambda \in [-0.37,-0.78]$.  

It is well known that exclusive $A(e,e'p)$ reactions, populating low-lying states in the 
residual $(A-1)^{*}$ nucleus, are proportional
to the FSI corrected single-particle momentum distributions for specific hole
states. Along similar lines, the SRC-driven $A(e,e'pN)$ cross section
is proportional to the c.m.~distribution of close-proximity pairs.
We find that the FSI only modestly affect the shape of the c.m. distribution, with the width of the distribution barely changing.
In essence, to a reasonable degree of accuracy, the
aggregated effect of FSI for exclusive $A(e,e'pN)$ processes is
a sizeable reduction of the plane-wave cross sections. This is a
remarkable result that could help to quantify the effect of
two-nucleon knockout contributions to quasi-elastic neutrino-nucleus
and anti-neutrino nucleus responses~\cite{neutrinoHammer}.

\begin{acknowledgments}
This work is supported by the
Research Foundation Flanders (FWO-Flanders) and by the Interuniversity Attraction Poles
Programme P7/12 initiated by the Belgian Science Policy Office. The computational
resources (Stevin Supercomputer Infrastructure) and services used in this work were provided
by Ghent University, the Hercules Foundation and the Flemish Government.
\end{acknowledgments}

\appendix 

\bibliography{twonucleonfsi}

\section{Qualitative model for the mass dependence of nuclear transparencies}
\label{app:toy_model}
In order to gain a qualitative understanding of the mass dependence of
the transparency in knockout reactions we develop a toy model. In
Sec.~\ref{subsec:toy_model} we introduce several scenarios each of which provides predictions for the mass dependence of the
nuclear transparency. We display the numerical results in Sec.~\ref{subsec:toy_results}.
\subsection{Model}
\label{subsec:toy_model}
We model the nucleus as a homogeneous sphere with radius $R = 1.20
A^{\frac{1}{3}} $\,fm and constant density $\rho(\vec{r})=
0.138~\text{fm}^{-3}$.  Without attenuation the $A(e,e'N)$ cross section is
proportional to the integrated density $\int \textrm{d}^{3} \vec{r}
\rho(\vec{r}) = A$.  The attenuation with the nuclear
medium is calculated with the aid of a classical survival probability $P(\vec{r}\,)$.
Given a nucleon brought into an energy continuum state at the coordinate
$\vec{r}$, $P(\vec{r}\,)$ is
\begin{align}
	P(\vec{r}\,) = \exp \left[ - \sigma \int_{z}^{+\infty} \textrm{d} z' \rho(\vec{r}\,') \right] \, .
	\label{eq:singleN_scat_prob}
\end{align}
Let $\vec{r} = (x,y,z)$ and $\vec{r}\,' = (x,y,z')$. The integration variable $z'$ runs along the direction of the momentum $\vec{p}$ of the outgoing nucleon. The cross section describing the scattering of the outgoing nucleon with the nuclear medium is denoted by $\sigma$. It is a measure for the aggregated effect of the attenuation.
For a sphere with radius $R$ and homogeneous density $\rho$ the survival probability of Eq.~(\ref{eq:singleN_scat_prob}) becomes,
\begin{align*}
	P(\vec{r}\,) = \exp\left[ - \sigma \rho ( \sqrt{ R^{2} - r^{2} \sin^{2} \xi } - r \cos^{2} \xi )\right] \, .
\end{align*}
Here, $\xi$ is the angle between $\vec{r}$ and $\vec{e}_{p} = \frac{\vec{p}}{p}$.
The $A(e,e'N)$ nuclear transparency $T^{N}_{A}$, defined as the cross section including attenuation divided by the cross section without attenuation, is
\begin{align}
	T^{N}_{A}[\text{single}] \propto \frac{\int \textrm{d}^{2} \Omega_{p} \int \textrm{d}^{3} \vec{r} \rho(\vec{r}) P(\vec{r})}{\int \textrm{d}^{2} \Omega_{p} \int \textrm{d}^{3} \vec{r} \rho(\vec{r})} = \frac{ 8 \pi^{2} \rho \int_{0}^{R} \textrm{d} r \, r^{2} \int_{-1}^{1} \textrm{d}x \exp\left[ - \sigma \rho ( \sqrt{ R^{2} - r^{2} (1-x^{2}) } - rx ) \right]}{ 4 \pi A } \, .
	\label{eq:transp_single}
\end{align}
The integration $\int \textrm{d}^{2} \Omega_{p}$ covers all possible outgoing-momentum directions.

%
%

For uncorrelated two-nucleon knockout the cross section is proportional to the total number of pairs $ \int \textrm{d}^{3} \vec{r}_1 \rho(\vec{r}_1)$ $\int \textrm{d}^{3} \vec{r}_2 \rho(\vec{r}_2) = A^{2}$.
The attenuation-corrected cross section is obtained by including the survival probability for both nucleons, and one finds for the two-nucleon knockout transparency $T_{A}^{NN}[\text{double}]$,
\begin{align}
	T_{A}^{NN}[\text{double}] \propto  \frac{ \int \textrm{d}^{2} \Omega_{p_1} \int \textrm{d}^{3} \vec{r}_1 \rho(\vec{r}_1) P(\vec{r}_1) \int \textrm{d}^{2} \Omega_{p_2} \int \textrm{d}^{3} \vec{r}_2 \rho(\vec{r}_2) P(\vec{r}_2)}{ \int \textrm{d}^{2} \Omega_{p_1} \int \textrm{d}^{3} \vec{r}_1 \rho(\vec{r}_1) \int \textrm{d}^{2} \Omega_{p_2} \int \textrm{d}^{3} \vec{r}_2 \rho(\vec{r}_2)} = T_{A}^{N}[\text{single}] \cdot T_{A}^{N}[\text{single}]
	\label{eq:transp_double}
\end{align}

Next we investigate two-nucleon knockout in the ZRA, which serves as a proxy for identifying SRC-prone nucleon pairs.
The ZRA is introduced by requiring that the initial nucleons are found at the same spatial coordinate.
The cross section without attenuation is proportional to,
\begin{align}
	\int  \textrm{d}^{2}\Omega_{p_1} \int \textrm{d}^{2}\Omega_{p_2} \int \textrm{d}^{3} \vec{r}_1 \rho(\vec{r}_1) \int \textrm{d}^{3} \vec{r}_2 \rho(\vec{r}_2) \delta(\vec{r}_1-\vec{r}_2) = (4 \pi)^{2} \int \textrm{d}^{3} \vec{r} \rho(\vec{r})^{2} = (4 \pi)^{3}  \rho^{2} \int_{0}^{R} \textrm{d} r \, r^{2}  = (4 \pi)^{2} \rho A
	\label{eq:doubleN_ZRA_PW}
\end{align}
We find that in the ZRA the two-nucleon knockout cross section is proportional to $A$ as opposed to $A^{2}$ in the uncorrelated case. Including attenuation gives,
\begin{multline}
	\int \textrm{d}^{3} \vec{r} \rho(\vec{r})^{2} \int \textrm{d}^{2} \Omega_{p_1} P_{1}(\vec{r}) \int \textrm{d}^{2} \Omega_{p_2} P_{2}(\vec{r}\,) = \\
	 16 \pi^{3} \rho^{2} \int_{0}^{R} \textrm{d} r \, r^{2}
	 \int_{-1}^{1} \textrm{d}x \exp\left[ - \sigma \rho ( \sqrt{ R^{2} - r^{2} (1-x^{2}) } - rx ) \right]  \int_{-1}^{1} \textrm{d}y \exp\left[ - \sigma \rho ( \sqrt{ R^{2} - r^{2} (1-y^{2}) } - ry ) \right] 
	\label{eq:doubleN_ZRA_FSI}
\end{multline}
The transparency mass dependence is then given by the ratio of Eqs.~(\ref{eq:doubleN_ZRA_FSI}) and~(\ref{eq:doubleN_ZRA_PW}),
\begin{align}
	T_{A}^{NN}[\text{ZRA}] \propto \frac{ \pi \rho}{A} \int_{0}^{R} \textrm{d} r \, r^{2}
	 \int_{-1}^{1} \textrm{d}x \exp\left[ - \sigma \rho ( \sqrt{ R^{2} - r^{2} (1-x^{2}) } - rx ) \right]  \int_{-1}^{1} \textrm{d}y \exp\left[ - \sigma \rho ( \sqrt{ R^{2} - r^{2} (1-y^{2}) } - ry ) \right] 
	 \label{eq:transp_doubleZRA}
\end{align}

%
%

It is well established that SRC pairs prefer back-to-back motion with anti-parallel momenta of the initial nucleon pair~\cite{Tang:2002ww,Colle:massdep}.
After introducing the following angular constraints in the ZRA cross sections of Eqs.~(\ref{eq:doubleN_ZRA_PW}),(\ref{eq:doubleN_ZRA_FSI}),
\begin{align*}
	\delta( \phi_1 - \phi_2 + \pi) \delta( \theta_1 + \theta_2 - \pi) \, ,
\end{align*}
the transparency becomes,
\begin{align}
	T_{A}^{NN}[\text{ZRA+SRC}] \propto & \frac{ 2 \pi \rho}{A} \int_{0}^{R} \textrm{d} r \, r^{2} \int_{-1}^{1} \textrm{d}\cos \theta_1 \exp\left[ - \sigma \rho ( \sqrt{ R^{2} - r^{2} (1-\cos^{2} \theta_1) } - r \cos \theta_1 ) \right]  \\
	& \times \int_{-1}^{1} \textrm{d}\cos \theta_2 \exp\left[ - \sigma \rho ( \sqrt{ R^{2} - r^{2} (1-\cos^{2} \theta_2) } - r \cos \theta_2 ) \right] \delta( \theta_1 + \theta_2 - \pi) \, .
\end{align}
With the substitution $(r,\cos \theta_1) \rightarrow (r,\ell = \sqrt{ R^{2} - r^{2} (1-\cos^{2} \theta_1) })$, and further manipulations one finds,
\begin{align}
	T_{A}^{NN}[\text{ZRA+SRC}] \propto &  \frac{ 2 \pi \rho}{A} \int_{0}^{R} \textrm{d} \ell \, \ell \exp( - 2 \rho \sigma \ell ) \sqrt{R^2-\ell^2} \ln \left( \frac{R+\ell}{R-\ell} \right) \, .
	\label{eq:transp_doubleZRA_SRC}
\end{align}
\subsection{Results}
\label{subsec:toy_results}
\begin{figure}
\centering
\includegraphics[width=0.8\textwidth]{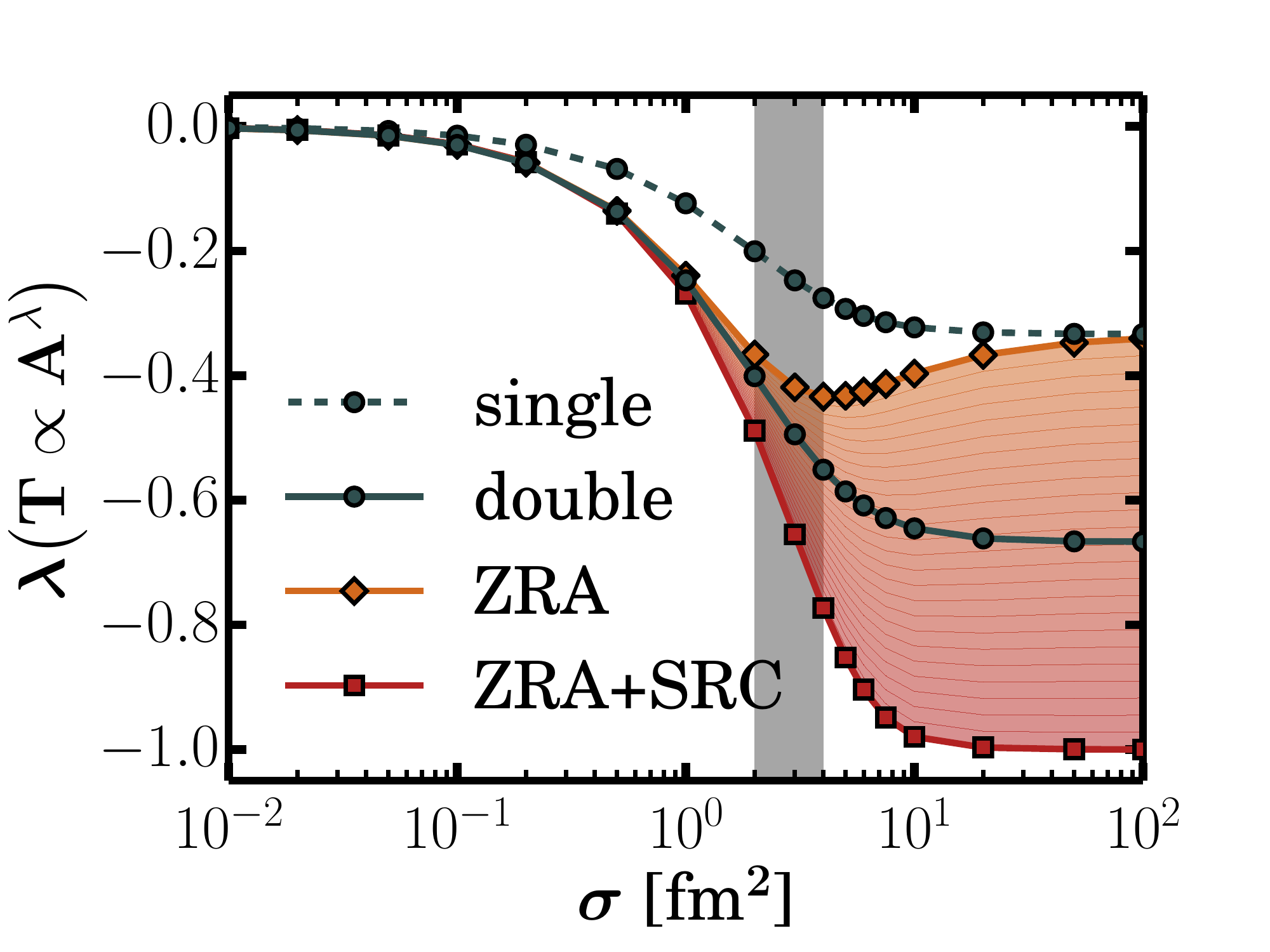}
\caption{(Color online) The exponents $\lambda$ in $T_{A}^{N(N)} \propto A^{\lambda}$ as a function of the nucleon-nucleus cross section $\sigma$.
The gray band corresponds with the $\sigma$ of outgoing nucleon momenta $0.3 \leq p \leq 10$ GeV/c. With ``single'' we denote the $T_{A}^{N}[\text{single}]$ results of Eq.~(\ref{eq:transp_single}). With ``double'' we refer to the $T_{A}^{NN}[\text{double}]$ results obtained with Eq.~(\ref{eq:transp_double}) which corresponds to uncorrelated two-nucleon knockout. The ``ZRA'' (``ZRA+SRC'') results for $T_{A}^{NN}[\text{ZRA}]$ ($T_{A}^{NN}[\text{ZRA+SRC}]$) are obtained with the expressions of Eq.~(\ref{eq:transp_doubleZRA})[Eq.~(\ref{eq:transp_doubleZRA_SRC})].}
\label{fig:T_dep}
\end{figure}

The mass dependence of the $T_{A}^{N},T_{A}^{NN}$ of Eqs.~(\ref{eq:transp_single}),(\ref{eq:transp_double}),(\ref{eq:transp_doubleZRA}),(\ref{eq:transp_doubleZRA_SRC}) is investigated by varying the mass number $A$ in the range $[12,208]$. A power law is fitted to the numerical results $T_{A} \propto A^{\lambda}$.
Figure~\ref{fig:T_dep} displays the numerical results for the exponent $\lambda$ as a function of $\sigma$.
In the limit of vanishing attenuation ($\sigma \rightarrow 0$) the cross section equals the plane-wave one and one has $T_{A} \approx A^{0}$.

For $\sigma > 10 \text{ fm}^{2}$ we find that the $\lambda$ values approach a limit value corresponding with an extremely opaque nucleus.
In this limit one expects that the single-nucleon knockout cross section is surface dominated $\propto A^{\frac{2}{3}}$ as no nucleons originating from within the nucleus are able to escape. The mass dependence of $T_{A}^{N}[\text{single}]$ then becomes,
\begin{align}
	\lim_{\sigma \rightarrow +\infty} T_{A}^{N}[\text{single}] \propto \frac{A^{\frac{2}{3}}}{A} = A^{-\frac{1}{3}} \, ,
\end{align}
which complies with the measured value \cite{Hen:2012jn,Lava:2004}.
For $T_{A}^{NN}[\text{double}]$, we have,
\begin{align}
 \lim_{\sigma \rightarrow +\infty} T_{A}^{NN}[\text{double}] = \lim_{\sigma \rightarrow +\infty} T_{A}^{N}[\text{single}] \cdot T_{A}^{N}[\text{single}] \propto A^{-\frac{2}{3}} \, .
\end{align}
In the case of two-nucleon knockout in the ZRA the two nucleons originate from the same spatial coordinate. We again expect a surface dominated cross section as in the single nucleon knockout case, leading to the exponent $\lim_{\sigma \rightarrow +\infty}  \lambda = -\frac{1}{3}$.

Including the additional constraint of back-to-back angles (ZRA+SRC in Fig.~\ref{fig:T_dep}) will strongly favour the situation in which $(\vec{r} \perp \vec{e}_{p_1} \Leftrightarrow \vec{r} \perp \vec{e}_{p_2}) \wedge (r \approx R)$. When investigating the mass dependence of the transparency in the strong attenuation limit one finds (Fig.~\ref{fig:T_dep}),
\begin{align}
	T_{A}^{NN}[\text{ZRA+SRC}] \propto \frac{A^{0}}{A^{1}} = A^{-1} \, .
\end{align}
The cross section in the strong attenuation limit becomes \textbf{independent} of $A$.

\end{document}